\documentclass[prd,eqsecnum,twocolumn,amsfonts,amssymb]{revtex4}

\usepackage{graphicx}

\usepackage{bm}

\newcommand{\beq}{\begin{equation}}
\newcommand{\eeq}{\end{equation}}
\newcommand{\beqs}{\begin{eqnarray}}
\newcommand{\eeqs}{\end{eqnarray}}

\begin{document}

\title{Study of Possible Ultraviolet Zero of the Beta Function in Gauge
Theories with Many Fermions}

\author{Robert Shrock}

\affiliation{C. N. Yang Institute for Theoretical Physics and \\
Department of Physics and Astronomy \\
Stony Brook University, Stony Brook, NY 11794 }

\begin{abstract}

We study the possibility of an ultraviolet (UV) zero in the $n$-loop beta
function of U(1) and non-Abelian gauge theories with $N_f$ fermions for large
$N_f$.  The effect of scheme transformations on the coefficients of different
powers of $N_f$ in the $n$-loop term in the beta function is calculated.  A
general criterion is given for determining whether or not the $n$-loop beta
function has a UV zero for large $N_f$.  We compare the results with exact
integral representations of the leading terms in the beta functions for the
respective Abelian and non-Abelian theories in the limit $N_f \to \infty$ limit
with $N_f \alpha$ finite. As part of this study, new analytic and numerical
results are presented for certain coefficients, denoted $b_{n,n-1}$, that
control the large-$N_f$ behavior at $n$-loop order in the beta function.  We
also investigate various test functions incorporating a power-law and essential
UV zero in the beta function and determine their manifestations in series
expansions in powers of coupling and in powers of $1/N_f$.

\end{abstract}


\maketitle


\section{Introduction}

The dependence of the running coupling constant in a (four-dimensional,
zero-temperature) gauge theory on the Euclidean momentum scale $\mu$ is of
fundamental field-theoretic interest. This dependence is determined by the beta
function of the theory \cite{rg}.  In this paper we will consider a U(1) gauge
theory with $N_f$ fermions of a given charge and a non-Abelian gauge theory
with gauge group $G$ and $N_f$ fermions transforming according to a
representation $R$ of $G$, in the limit of large $N_f$.  The fermions are
assumed to have masses that are zero or negligibly small relative to the
relevant range of scales $\mu$.  Both of these theories have positive beta
functions for small gauge coupling where they are perturbatively calculable, so
they are infrared-free in this region of coupling.  

We present several new results here.  First, we investigate the question of a
possible ultraviolet zero in a U(1) gauge theory further, extending our recent
study in \cite{sch2}.  We analyze zeros of the $n$-loop $\beta$ function as a
function of $\alpha$ for a large range of $N_f$.  We use our computations to
test and confirm an approximate analytic solution to the equation for a UV zero
of the $n$-loop $\beta$ function.  Using the results of \cite{sch2,sch}, we
calculate the effect of scheme transformations on the coefficients of different
powers of $N_f$ in the $n$-loop term in the beta function for U(1) and
non-Abelian gauge theories.  We deduce a general criterion for the existence of
a UV zero in the $n$-loop $\beta$ function for large $N_f$ applicable for both
Abelian and non-Abelian gauge theories.  We compare the results with
implications from an exact integral representation for the leading term,
denoted $F_1$, in the beta function of a U(1) theory at large-$N_f$.  Up to
loop order $n=24$ to which we probe, we do not find evidence for a stable UV
zero in the U(1) beta function that would be reached from small coupling as
$\mu$ is increased. We provide some insight into this by showing that the
coefficients, denoted $b_{n,n-1}$, of the leading part of the $n$-loop term,
$b_n$, in the beta function show a scatter of positive and negative signs,
while the $b_{n,n-1}^{(d)}$ that correspond to the part of $F_1$ that might be
responsible for a UV zero are all negative. Additional insight is obtained from
calculations of series expansions obtained from various illustrative test
functions for $\beta$.  We also carry out a similar analysis for a non-Abelian
gauge theory with fermions in various representations, in the large-$N_f$
limit, reaching a similar conclusion.

We recall some relevant background and previous work. If a beta function of a
quantum field theory is positive near zero coupling, the coupling grows as
$\mu$ increases.  However, the beta function may have an ultraviolet zero, so
that the coupling approaches a constant as $\mu \to \infty$. An explicit
example of this occurs in the O($N$) nonlinear $\sigma$ model in $d=2+\epsilon$
spacetime dimensions.  From an exact solution of this model in the limit $N \to
\infty$ (involving a sum of an infinite number of Feynman diagrams that
dominate in this limit), one finds that the beta function for small $\epsilon$
has the form \cite{nlsm}
\beq
\beta(\lambda) = \epsilon \lambda \Big ( 1 - \frac{\lambda}{\lambda_c} \Big ) 
\ , 
\label{betanlsm}
\eeq
where $\lambda$ is the effective coupling and $\lambda_c = 2\pi
\epsilon/N$. Thus, assuming that $\lambda$ is small for small $\mu$, it follows
that as $\mu$ increases, $\lambda$ approaches the UV fixed point
at $\lambda_c$ as $\mu \to \infty$.

It was observed early in the history of work on quantum electrodynamics (QED)
that the property that the beta function of QED is positive for small
couplings, where it is perturbatively calculable, implies that the theory is
free as $\mu \to 0$ in the infrared (IR) \cite{landau}.  The positive $\beta$
function means that as $\mu$ increases toward the ultraviolet, the gauge
coupling grows in strength.  Indeed, integrating the one-loop renormalization
group (RG) equation would yield a pole at finite $\mu$ in the ultraviolet, the
Landau pole.  Of course, one cannot reliably use the $\beta$ function,
perturbatively calculated to one-loop or even higher-loop order, for values of
$\mu$ where the coupling gets large, so there is no rigorous implication that
the theory would, in fact, have a Landau pole.  However, this led early
researchers to inquire whether the $\beta$ function of a U(1) gauge theory
might exhibit a UV zero away from the origin.  If such a UV zero of the beta
function could be demonstrated reliably, then as the Euclidean scale $\mu$
increased, the gauge coupling would approach a finite value rather than
continuing to increase, i.e., this would be a UV fixed point of the
renormalization group.  A necessary condition for the analysis to be reliable
would be that the UV zero must occur at a reasonably small value of the gauge
coupling. The calculation of the two-loop term in the $\beta$ function for this
theory \cite{b2u1} found that it was positive, like the one-loop term, and
hence excluded the existence of a UV zero of $\beta$ at this loop order. 

The approach to the analysis of a possible UV zero in the beta function for the
electromagnetic U(1)$_{em}$ gauge theory changed after the development of the
${\rm SU}(2)_L \otimes {\rm U}(1)_Y$ electroweak sector of the Standard Model
(SM), since in the SM, the photon field $A_\mu^{(\gamma)}$ arises upon
electroweak symmetry breaking as the linear combination $A_\mu^{(\gamma)} =
\cos\theta_W \, B_\mu + \sin\theta_W \, A^3_\mu$, where ${\vec A}_\mu$ and
$B_\mu$ are the gauge bosons for the SU(2)$_L$ and weak hypercharge U(1)$_Y$
factor groups, and $\theta_W$ is the weak mixing angle.  Above the electroweak
symmetry breaking scale, and hence for considerations of asymptotic ultraviolet
behavior, the Abelian gauge interaction that one naturally analyzes is
U(1)$_Y$. Although this is chiral, as contrasted with the vectorial U(1)$_{em}$
interaction, it shares the property of being non-asymptotically free. Moreover,
if the U(1)$_Y$ gauge group is embedded in an asymptotically free simple gauge
group at some mass scale $M_{GUT}$, as in grand unified theories, then one need
not worry about the asymptotic behavior of the U(1)$_Y$ gauge interaction in
the UV at scales above $M_{GUT}$. 

Nevertheless, the renormalization-group properties of a vectorial U(1) gauge
theory have continued to be of abstract field-theoretic interest.  In
particular, higher-loop terms in the beta function of such a theory have been
calculated \cite{b3u1Nf1}-\cite{b5u1b}.  For our analysis of an Abelian gauge
theory, we will focus on this type of vectorial theory here.  A number of
studies of possible nonperturbative properties of a U(1) gauge theory have been
carried out over the years using approximate solutions of Schwinger-Dyson
equations and other methods \cite{jbw}-\cite{gies04}). In particular, lattice
studies of U(1) gauge theories (with dynamical staggered fermions) have been
performed in \cite{scgt90,schierholz,kogut}. A useful result is an exact
calculation of the coefficient, $F_1(y)$, of the leading $1/N_f$ correction to
the $\beta$ function in the large-$N_f$ limit \cite{pascual}, where $y$ is
proportional to a product of $N_f$ times the squared gauge coupling.  An
analogous exact large-$N_f$ calculation for a non-Abelian gauge theory was
given in \cite{graceybeta} (related exact large-$N_f$ results for the anomalous
dimension of the fermion bilinear were reported in \cite{graceygamma}).  

An interesting analysis using the exact $N_f \to \infty$ results for the beta
functions of the U(1) and non-Abelian gauge theories to explore possible zeros
of the beta function has been carried out by Holdom in \cite{holdom2010} (see
also \cite{ps}). Holdom observed that $F_1(y)$ diverges logarithmically through
negative values as $y$ approaches a certain value ($y=15/8$) from below, and
hence the beta function, calculated to order $1/N_f$, which is proportional to
$(1+F_1(y)/N_f)$, has a UV zero. However, as he noted, terms of higher order in
$1/N_f$ could modify this, so that the beta function might very well not, in
fact, have a UV zero.

Most studies of the renormalization-group behavior of non-Abelian gauge
theories were motivated by the approximate Bjorken scaling observed in deep
inelastic scattering and the property of asymptotic freedom that explains this,
as part of the theory of quantum chromodynamics (QCD) \cite{b1}.  The beta
function for a vectorial non-Abelian gauge theory has been calculated up to
four-loop order \cite{b1}-\cite{gross75}.  A convenient scheme which has been
widely used is the $\overline{\rm MS}$ scheme \cite{ms,msbar}. Calculating
higher-order terms in the beta function of a gauge theory to three-loop and
higher-loop level is useful because the results give a quantitative measure of
the accuracy of the two-loop calculation.  Indeed, in QCD, the value of the
higher-loop calculations has been amply demonstrated by their use in fitting
data on the $Q^2=\mu^2$ dependence of the strong coupling $\alpha_s(\mu)$
\cite{bethke}. 

 For a given non-Abelian gauge group $G$ and fermion representation $R$, if the
number, $N_f$, of fermions is small, the theory is confining, with spontaneous
chiral symmetry breaking.  As $N_f$ increases further, the theory 
exhibits an approximate or exact infrared zero in the beta function
\cite{b2,bz}.  For sufficiently large $N_f$, this occurs at a small value of
the coupling, so that one expects the theory to evolve from the ultraviolet to
a deconfined, chirally symmetric Coulombic phase in the infrared.  This IR zero
has been studied at higher-loop level in \cite{gk98}-\cite{lnn}. Effects of
scheme transformations on the position of the IR zero have been investigated
recently in \cite{sch,sch2,rc} (see also \cite{gracey4loop,pqcd}). These can
also be applied to the analysis of a UV zero in the beta function. As $N_f$
increases sufficiently, the sign of the leading, one-loop term is reversed, and
the theory becomes infrared-free rather than ultraviolet-free.  In this regime,
one can then examine the non-Abelian theory for a possible UV zero in the beta
function. Since in this regime (as reviewed below) the one-loop and two-loop
terms have the same sign, the beta function does not have such UV zero at the
maximal scheme-independent level for general $N_f$.  This is the same situation
as for the U(1) theory, and as in the Abelian case, one may then investigate
higher-loop terms in the beta function to see if such a UV zero might appear.
Furthermore, one may study how such a UV zero, if present, relates to the exact
results in the $N_f \to \infty$ limit. We address this question here for both
U(1) and non-Abelian gauge theories.

 This paper is organized as follows.  In Sect. \ref{general} we investigate a
possible UV zero in the $n$-loop beta function of a U(1) gauge theory. In
Sect. \ref{u1uvzero} we discuss the general structure of the beta function and
calculate the effect of a scheme transformation on the coefficients of the
various powers of $N_f$ in the $n$-loop term.  Sect. \ref{u1uvzero} presents
our analysis of a possible UV zero in the $n$-loop beta function, in
particular, for large $N_f$. In Sect. \ref{u1betalnf} the results of this
analysis are compared with implications from an exact calculation of the
coefficient of the leading $1/N_f$ correction term in an appropriately rescaled
beta function in the large-$N_f$ limit. In Sect. \ref{testfunctions} we carry
out a study of various test functions incorporating a UV zero in the beta
function, of both a power-law and essential-zero form, to determine the
manifestations that they produce in both kind of series expansions, namely an
expansion in small $1/N_f$ for fixed $N_f \alpha$ and an expansion in $\alpha$
for fixed large $N_f$. Sect. \ref{nagt} contains corresponding results for a
non-Abelian gauge theory.  Our conclusions are given in 
Sect. \ref{conclusions}.  Some relevant formulas are given in appendices
\ref{nfstappendix}-\ref{rcvalues}.


\section{UV Zero of the $n$-loop Beta Function of a Gauge Theory} 
\label{general}


\subsection{General} 

In this section we discuss some general features of the beta function for a
gauge theory.  The discussion in the first subsection applies to both an
Abelian and a non-Abelian gauge theory; we comment on relevant differences at
appropriate points subsequently. The Abelian U(1) theory contains $N_f$
fermions of a given charge $q$, while the non-Abelian theory has $N_f$ fermions
transforming according to a representation $R$ of the gauge group $G$.  The
fermions are assumed to have masses that are negligibly small relative to the
Euclidean momentum scale, $\mu$.  We denote the running gauge coupling as
$g(\mu)$. In addition to the standard notation $\alpha(\mu) = g(\mu)^2/(4\pi)$,
it will be convenient to use the quantity
\beq
a(\mu) = \frac{g(\mu)^2}{16 \pi^2} = \frac{\alpha(\mu)}{4\pi} \ . 
\label{a}
\eeq
The scale $\mu$ will often be suppressed in the notation.  In the Abelian case,
with no loss of generality, we absorb the factor $q$ into a rescaling of the
coupling $g$ and hence set $q=1$.  The dependence of $\alpha$ on $\mu$ is given
by the $\beta$ function
\beq
\beta \equiv \beta_\alpha \equiv \frac{d\alpha}{dt} \ , 
\label{betadef}
\eeq
where $dt=d \ln \mu$. This has the series expansion \cite{signconvention} 
\beq
\beta_\alpha = 2\alpha \sum_{\ell=1}^\infty b_\ell \, a^\ell =
 2\alpha \sum_{\ell=1}^\infty \bar b_\ell \, \alpha^\ell \ ,
\label{beta}
\eeq
where $\ell$ denotes the number of loops involved in the calculation of
$b_\ell$ and $\bar b_\ell = b_\ell/(4\pi)^\ell$.  The one-loop and two-loop
coefficients $b_1$ and $b_2$ in the beta function are independent of the scheme
used for regularization and renormalization, while the coefficients at
higher-loop order $n \ge 3$ are scheme-dependent \cite{gross75}. The $n$-loop
($n\ell$) $\beta$ function, denoted $\beta_{\alpha,n\ell}$, is defined by
(\ref{beta}) with the upper limit on the summation over loop order $\ell$ given
by $\ell=n$ rather than $n=\infty$.

The U(1) gauge theory is infrared-free and, for the regime of large $N_f$ in
which we are interested, the non-Abelian theory is also infrared-free. A zero
of the beta function that is reached by renormalization-group evolution from
the vicinity of zero coupling is thus a UV zero.  The condition that the
$n$-loop $\beta$ function vanishes is the polynomial equation
\beq
\sum_{\ell=1}^n b_\ell a^{\ell-1} = 0 
\label{beta_nloop_zero}
\eeq
of degree $n-1$ in $a$ or equivalently, $\alpha$.  The $n-1$ roots of this
equation depend on $n-1$ ratios of the $n$ coefficients, which can be taken to
be $b_\ell/b_n$ with $\ell=1,...,n-1$.  The real positive root nearest to the
origin (if it exists) is the UV zero of $\beta_{\alpha,n\ell}$.  We will denote
this zero as $a_{_{UV,n\ell}} = \alpha_{_{UV,n\ell}}/(4\pi)$.  For later
purposes, let us assume that for a given $N_f$, $\beta_{\alpha,n\ell}$ has a
zero at $\alpha_{_{UV,n\ell}}$, and let us define the interval
$I_{\alpha,n\ell}$ as
\beq
I_{\alpha,n\ell}: \quad 0 \le \alpha \le \alpha_{_{UV,n\ell}} \ . 
\label{alphainterval}
\eeq
We will assume that $\alpha(\mu)$ is sufficiently small for small values of
$\mu$ in the IR that perturbation theory is reasonably reliable.  Then as $\mu$
increases from the IR to the UV, $\alpha(\mu)$ increases from these small
values and approaches the UV zero of $\beta_\alpha$ at
$\alpha_{_{UV,n\ell}}$. 

If a UV zero of $\beta_\alpha$ calculated at a given loop order occurs at a
value of $\alpha \sim O(1)$, it is important to take account of higher-loop
contributions to determine how they affect the position of this UV zero. We
performed such a study for U(1) with various values of $N_f$ in \cite{sch2}.
Here we generalize this to larger $N_f$ and investigate how these calculations
using a small-$\alpha$ expansion at fixed $N_f$ relate to results obtained as
$N_f \to \infty$ for fixed $N_f \alpha$. We will also explore a UV zero for
non-Abelian gauge theories at large $N_f$. To distinguish quantities in a U(1)
and a non-Abelian (NA) gauge theory, we will use the symbols $\beta_\alpha$,
$b_\ell$, etc. to refer to the U(1) theory and $\beta_\alpha^{(NA)}$,
$b_\ell^{(NA)}$, etc. to refer to the non-Abelian theory.


\subsection{Effect of Scheme Transformations on $b_n$} 
\label{nfst}

We discuss here the structure of the $n$-loop terms in the U(1) beta function
as polynomials in $N_f$ and calculate how this structure changes under a scheme
transformation.  The one-loop and two-loop coefficients in the beta function
are both proportional to $N_f$, and we write
\beq
b_1 = b_{1,1}N_f \ , \quad b_2 = b_{2,1}N_f \ . 
\label{b1b2form}
\eeq
where the values of $b_{\ell,1}$ are given in Eqs. (\ref{b1u1}) and
(\ref{b2u1}) below. Although the higher-loop coefficients, $b_\ell$ with $\ell
\ge 3$, are scheme-dependent, one can make some general statements about their
dependence on $N_f$ from the structure of the Feynman diagrams that contribute
to these higher-loop coefficients. For $\ell \ge 2$, $b_\ell$ is a polynomial
in $N_f$ in which the term of lowest degree in $N_f$ has degree 1 and the term
of highest degree in $N_f$ has degree $\ell-1$.  That is, these coefficients
have the structural form
\beq
b_\ell = \sum_{k=1}^{\ell-1} b_{\ell,k} \, N_f^k \quad {\rm for} \ \ell \ge 2
\ . 
\label{bellexpansion}
\eeq
As defined in this manner, the $b_{\ell,k}$ are independent of $N_f$. For 
later purposes, we will formally extend the range of the index $k$ in 
$b_{\ell,k}$ to allow $k=\ell$ and set $b_{\ell,\ell} = 0$ for $\ell \ge 2$. 

We next investigate the effect of scheme transformations on the $b_{\ell,k}$. 
A scheme transformation (ST) can be expressed as a mapping between
$\alpha$ and $\alpha'$, or equivalently, $a$ and $a'$:
\beq
a = a' f(a') \ .
\label{aap}
\eeq
To keep the UV properties the same, one requires $f(0) = 1$.  We consider
scheme transformations here that are analytic about $a=a'=0$ \cite{nonan}
and hence can be expanded in the form
\beq
f(a') = 1 + \sum_{s=1}^{s_{max}} k_s (a')^s =
        1 + \sum_{s=1}^{s_{max}} \bar k_s (\alpha')^s \ ,
\label{faprime}
\eeq
where the $k_s$ are constants, $\bar k_s = k_s/(4\pi)^s$, and $s_{max}$ may be
finite or infinite. The beta function in the transformed scheme is 
\beq
\beta_{\alpha'} \equiv \frac{d\alpha'}{dt} = \frac{d\alpha'}{d\alpha} \,
\frac{d\alpha}{dt} \ .
\label{betaap}
\eeq
This has the expansion
\beq
\beta_{\alpha'} = 2\alpha' \sum_{\ell=1}^\infty b_\ell' (a')^\ell =
2\alpha' \sum_{\ell=1}^\infty \bar b_\ell' (\alpha')^\ell \ ,
\label{betaprime}
\eeq
where $\bar b'_\ell = b'_\ell/(4\pi)^\ell$. Explicit expressions for the
$b'_\ell$ in terms of the $k_s$ determining the transformation function were
calculated in \cite{sch} and studied further in \cite{sch2}.  In our discussion
below on effects of scheme transformations on the coefficients $b_\ell$ to
produce the $b_\ell'$, we implicitly restrict to $b_\ell'$ with $\ell \ge 3$,
since there is no change in the $b_\ell$ for $\ell=1, \ 2$. A general property
that was found in \cite{sch} is that $b_\ell'$ is a linear combination of $b_n$
with $1 \le n \le \ell$, of the form
\beq
b_\ell' = b_\ell + \sum_{m=1}^{\ell-1} h_{\ell,\ell-m} b_{\ell-m} \ , 
\label{bprimerelation}
\eeq
where the $h_{\ell,n}$ are functions of the $k_s$ that determine the scheme
transformation, as given in Eq. (\ref{faprime}).  For example, for $\ell=3$ and
$\ell=4$, we calculated \cite{sch}
\beq
b_3' = b_3 + k_1b_2+(k_1^2-k_2)b_1 
\label{b3prime}
\eeq
and
\beq
b_4' = b_4 + 2k_1b_3+k_1^2b_2+(-2k_1^3+4k_1k_2-2k_3)b_1 \ , 
\label{b4prime}
\eeq
so that $h_{3,2}=k_1$, $h_{3,1} = k_1^2-k_2$, and so forth for higher
$\ell$. For the range $\ell \ge 3$ where the $b_\ell$ change under a
scheme transformation, $h_{\ell,\ell-1}=(\ell-2)k_1$. 

To determine the effect of a scheme transformation on the coefficients
$b_{\ell,k}$, we apply the results above.  A general scheme transformation
changes the $b_\ell$ for $\ell \ge 3$ and, in particular, the $b_{\ell,k}$ for
$\ell \ge 3$ and $1 \le k \le \ell-1$.  Since we are interested in a specific
limit, $N_f \to \infty$, we will restrict ourselves here to scheme
transformations that are independent of $N_f$.  (We will sometimes emphasize
this with the symbol NFI, standing for ``$N_f$-independent''.)  Substituting
Eq. (\ref{bellexpansion}) into Eq. (\ref{bprimerelation}), we obtain the result
that $b_\ell'$ again has the same general polynomial form in terms of powers of
$N_f$ as the $b_\ell$ in Eq. (\ref{bellexpansion}), viz.,
\beq
b_\ell' = \sum_{k=1}^{\ell-1} b_{\ell,k}' N_f^k \ . 
\label{bellprimeexpansion}
\eeq
Explicitly, we have, for the values $\ell \ge 3$ where the $b_\ell$ change
under a scheme transformation, 
\beq
\sum_{k=1}^{\ell-1} b_{\ell,k}' N_f^k =
\sum_{k=1}^{\ell-1} b_{\ell,k} N_f^k +
\sum_{n=1}^{\ell-1} h_{\ell,n} t_n 
\label{bellktransf}
\eeq
where $t_1=b_{1,1}N_f$ and 
\beq
t_n = \sum_{r=1}^{n-1} b_{n,r} N_f^r \quad {\rm if} \ 2 \le n \le \ell-1 \ . 
\label{tn}    
\eeq
The highest power of $N_f$ in the second term in Eq. (\ref{bellktransf}) is
$r=n-1=\ell-2$, so the only term of the form $N_f^{\ell-1}$ on the right-hand
side of Eq. (\ref{bellktransf}) arises from the first term and is
$b_{\ell,\ell-1}N_f^{\ell-1}$.  Therefore, 
\beq
b_{\ell,\ell-1}' = b_{\ell,\ell-1} \ . 
\label{bnnminus1invariance}
\eeq
Thus, although the $b_\ell$ coefficients with $\ell \ge 3$ that we use for our
calculations were calculated in the $\overline{MS}$ scheme, in each $b_\ell$,
the term with the highest power of $N_f$, namely $b_{\ell,\ell-1}N_f^{\ell-1}$,
is invariant under NFI scheme transformations. This has an important
consequence, namely that although each term $b_\ell$ with $\ell \ge 3$ is
scheme-dependent, the term in $b_\ell$ that dominates in the $N_f \to \infty$
limit is NFI scheme-independent. In effect, this limit picks out the
NFI scheme-independent part of the beta function to arbitrarily high loop order
$\ell$.  The same type of argument holds for a non-Abelian gauge theory, so
(for $\ell \ge 3$ where the $b_\ell^{(NA)}$ change under an NFI scheme
transformation)
\beq
b_{\ell,\ell-1}^{(NA) \prime} = b_{\ell,\ell-1}^{(NA)} \ . 
\label{bnnminus1invariance_nagt}
\eeq

Equating equal powers of $N_f$ on the left-hand and right-hand sides of
Eq. (\ref{bellktransf}), from $N_f^{\ell-2}$ down to $N_f$, we obtain a
relations between the $b_{\ell,k}'$ and $b_{\ell,k}$ involving the coefficients
in Eq. (\ref{bprimerelation}) with $1 \le k \le \ell-2$, namely
\beq
b_{\ell,k}' = b_{\ell,k}+\sum_{m=1}^{\ell-k-1} h_{\ell,\ell-m} \, b_{\ell-m,k} 
\ , 
\label{bellkprimebellk}
\eeq
and, for $k=1$, 
\beq
b_{\ell,1}' = b_{\ell,1}+\sum_{m=1}^{\ell-1} h_{\ell,\ell-m} \, b_{\ell-m,1} \
. 
\label{bell1primebell1}
\eeq
We will use these relations below. 


\section{UV Zero of the $n$-loop Beta Function of a U(1) Gauge Theory} 
\label{u1uvzero}

Having discussed the general structure of the beta function and the condition
for a zero in this function, we next proceed to our actual calculations for the
U(1) gauge theory.  The coefficients of the one-loop and two-loop terms of
$\beta_\alpha$ in this U(1) theory are \cite{rg,b2u1}
\beq
b_1 = \frac{4N_f}{3}
\label{b1u1}
\eeq
and
\beq
b_2 = 4N_f \ .
\label{b2u1}
\eeq
As noted above, since these coefficients have the same sign, the two-loop
$\beta_\alpha$ function, $\beta_{\alpha,2\ell}$, does not have a UV zero 
\cite{cgtu1}. For later purposes it will be convenient to extract the
coefficients of powers of $N_f$ in these $b_\ell$s, defining 
\beq
b_{1,1} \equiv \frac{b_1}{N_f} = \frac{4}{3} 
\label{b11}
\eeq
and
\beq
b_{2,1} \equiv \frac{b_2}{N_f} = 4 \ . 
\label{b21}
\eeq

The coefficient of the three-loop term in the $\beta_\alpha$ function of the
U(1) gauge theory, calculated in the $\overline{\rm MS}$ scheme, is
\cite{b3u1Nf1,b3u1}
\beq
b_3 = -2N_f\Big ( 1 + \frac{22N_f}{9} \Big ) \ .
\label{b3u1}
\eeq
This is evidently negative for all $N_f$.  Therefore, in addition to the
IR zero at $\alpha=0$, in the $\overline{\rm MS}$ scheme, the three-loop
$\beta_\alpha$ function, $\beta_{\alpha,3\ell}$, has a UV zero, namely, 

\beq
\alpha_{_{UV,3\ell}}= 4 \pi a_{_{UV,3\ell}} =
 \frac{4\pi[9 + \sqrt{3(45+44N_f)} \ ]}{9+22N_f} \ . 
\label{alfuv_3loop}
\eeq
In \cite{sch2} we calculated values of $\alpha_{_{UV,3\ell}}$ as a function
of $N_f$ from 1 to 10.  For our present large-$N_f$ study, we extend this to 
$N_f=10^4$ and present results in Table \ref{alfuvu1}. In 
addition to the scheme-dependence, one must note that for moderate $N_f$, the
value of $\alpha_{_{UV,3\ell}}$ in Eq. (\ref{alfuv_3loop}) is too large for
the perturbative three-loop calculation to be very reliable.  

The coefficient of the four-loop term in $\beta_\alpha$, again 
calculated in the $\overline{\rm MS}$ scheme, is \cite{b4u1} 
\beq
b_4 = N_f\Big [ -46 + \Big ( \frac{760}{27} - \frac{832\zeta(3)}{9} \Big ) 
N_f - \frac{1232}{243}N_f^2 \Big ] \ , 
\label{b4u1}
\eeq
where $\zeta(s) = \sum_{n=1}^\infty n^{-s}$ is the Riemann zeta function.
Numerically,
\beq
b_4 = -N_f \, [46+82.97533N_f+5.06996N_f^2] \ .
\label{b4}
\eeq
As was the case with $b_3$, the coefficient $b_4$ is negative for all $N_f >
0$. The condition that $\beta_{\alpha,4\ell}=0$ for $\alpha \ne 0$, is the
cubic equation $b_1+b_2a+b_3a^2+b_4a^3=0$.  This equation has a physical root,
$a_{_{UV,4\ell}}=\alpha_{_{UV,4\ell}}/(4\pi)$, as well as an unphysical pair of
complex-conjugate values of $a$.  We list values of $\alpha_{_{UV,4\ell}}$ for
$N_f$ up to $10^4$ in Table \ref{alfuvu1}.  As was true of
$\alpha_{_{UV,3\ell}}$, in this $\overline{\rm MS}$ scheme,
$\alpha_{_{UV,4\ell}}$ is a monotonically decreasing function of $N_f$.

The coefficient of the five-loop term in $\beta_\alpha$, $b_5$, is, in 
the $\overline{\rm MS}$ scheme \cite{b5u1a,b5u1b}
\begin{widetext}
\beqs
b_5 & = & N_f \bigg [ \frac{4157}{6}+128\zeta(3)
+ \Big (-\frac{7462}{9}-992\zeta(3)+2720\zeta(5) \Big ) N_f \cr\cr
& + & \Big (-\frac{21758}{81}+\frac{16000\zeta(3)}{27}-\frac{208\pi^4}{135}
-\frac{1280\zeta(5)}{3} \Big )N_f^2 +
\Big (\frac{856}{243}+\frac{128\zeta(3)}{27} \Big )N_f^3 \bigg ] \ .
\label{b5u1}
\eeqs
\end{widetext}
Note that here and below, terms involving $\zeta(m)$ with even $m=2r$ are
evaluated using the identity (\ref{zetaeven}) in Appendix \ref{gammafunction}.
Numerically, 
\beqs
b_5 & = & N_f( 846.6966 + 798.8919N_f - 148.7919N_f^2 
\cr\cr
& + & 9.22127N_f^3 \ . 
\label{b5num}
\eeqs
This is positive for all non-negative $N_f$.  The condition that
$\beta_{\alpha,5\ell}$ vanishes away from the origin is given by
Eq. (\ref{beta_nloop_zero} with $n=5$, which is a quartic equation in $a$. In
\cite{sch2} we calculated this for $1 \le N_f \le 10$ (see Table II of
\cite{sch2}). For $1 \le N_f \le 4$, $\beta_{\alpha,5\ell}$ has no UV zero.
For $5 \le N_f \le 19$, $\beta_{\alpha,5\ell}$ does have a UV zero, but as
$N_f$ (analytically continued to real values \cite{nfintegral}) increases
through $N_f \simeq 19.67$, this zero disappears, and for larger values of
$N_f$, the roots of the quartic equation above consist of two complex-conjugate
pairs.

In general, insofar as a perturbative expansion of a given quantity in a field
theory is reliable, one would expect that calculating this quantity to
higher-loop order should not change its qualitative properties.  Indeed, one
would expect that the fractional change in the quantity going from $n$ to $n+1$
loops should become progressively smaller as the loop order $n$ increases.
According to this expectation, having calculated the UV zero in $\beta$ at
three-loop and four-loop order, one would expect that the result of a five-loop
calculation would be a small shift in the zero of $\beta$. As is evident from
our results in Table \ref{alfuvu1}, we do not, in general, find that the UV
zeros that we have calculated at $n=3$, $n=4$, and $n=5$ loop order for a large
range of $N_f$ values satisfy this necessary condition.  In the small interval
of $N_f$ from about 6 to roughly 10, the behavior of the UV zero is reasonably
stable.  But for larger $N_f$, the behavior is not stable. Indeed, for $N_f \ge
20$, the five-loop beta function (calculated in the $\overline{\rm MS}$ scheme)
does not even have a UV zero.  Thus, as far as it has been calculated, the
perturbative beta function of this U(1) gauge theory does not exhibit a stable
UV zero for large $N_f$.  Since this is a statement about the perturbative beta
function calculated to a finite-loop level, it does not contradict the possible
existence of a UV zero of the beta function in the limit $N_f \to \infty$
obtained from a summation over a leading set of diagrams up to infinite-loop
order.  We will discuss this $N_f \to \infty$ limit next.


\section{Beta Function of the U(1) Gauge Theory in the Large-$N_f$ Limit} 
\label{u1betalnf}

\subsection{Large-$N_f$ Limit and Rescaled Beta Function $\beta_y$} 

For a U(1) gauge theory with $N_f$ fermions, we consider the limit 
\beq
N_f \to \infty \quad {\rm with} \quad y(\mu) \equiv N_f \, a(\mu) = 
\frac{N_f \, \alpha(\mu)}{4\pi} \ , 
\label{lnf}
\eeq
where the function $y(\mu)$ is a finite function of the Euclidean scale $\mu$
in the $N_f \to \infty$.  We denote this as the LNF (large-$N_f$) limit. We
will use the same term, LNF, below for the corresponding limit 
(\ref{lnf_nagt}) in the case of a non-Abelian gauge theory. 
For notational simplicity, we will often suppress the
argument $\mu$ and write $y(\mu)$ as $y$.  The LNF limit is useful because 
certain Feynman diagrams, namely iterated fermion vacuum polarization
insertions on the Abelian gauge boson propagator lines, give the dominant
contribution to $\beta$ in this limit, and this contribution can be calculated
exactly in terms of an integral representation, to be discussed below.
Although for definiteness we phrase our discussion of the $N_f \to \infty$
limit here in terms of the U(1) gauge theory, it will also apply, with obvious 
changes, to the non-Abelian gauge theory to be analyzed later in the paper. 

From the structural formula (\ref{bellexpansion}), it follows that 
\beq 
b_\ell \propto N_f^{\ell-1} \quad {\rm for} \ \ \ell \ge 2 \quad {\rm as}
\ \ N_f \to \infty \ .
\label{bellnf}
\eeq

The asymptotic relation (\ref{bellnf}) may be contrasted with
the dependence of the $b_\ell$ coefficients in an SU($N_c$) theory in the
large-$N_c$ limit, which is
\beq
b_\ell \propto N_c^\ell \quad \forall \ \ell \quad {\rm as} \ N_c \to \infty 
\ . 
\label{bellnc}
\eeq
The large-$N_f$ dependence of the coefficients $b_\ell$ in $\beta_\alpha$
motivates the definition of rescaled coefficients that have finite limits as
$N_f \to \infty$, namely $b_1/N_f = b_{1,1} = 4/3$, as given above, and 
\beq
\check b_\ell \equiv \frac{b_\ell}{N_f^{\ell-1}} \quad {\rm for} \ \ell \ge 2
\ . 
\label{bellck}
\eeq
Combining these definitions with Eq. (\ref{bellexpansion}), we have
\beq
\lim_{N_f \to \infty} \check b_\ell = b_{\ell,\ell-1} \quad {\rm for} \ 
\ell \ge 2 \ . 
\label{bellleading}
\eeq
For compact notation, we also define
\beq
\acute b_\ell \equiv \frac{\check b_\ell}{b_{1,1}} = \frac{3}{4}\check b_\ell
\label{bellacute}
\eeq
and
\beq
\acute b_{\ell,k} \equiv \frac{b_{\ell,k}}{b_{1,1}} = \frac{3}{4}b_{\ell,k} 
\ . 
\eeq

Given that $b_1 \propto N_f$ and the structural relation 
(\ref{bellexpansion}) for the $b_\ell$ with $\ell \ge 2$, one may construct a
rescaled beta function $\beta_y$ that is finite in the LNF limit by defining 
\beq
\beta_y \equiv \beta_\alpha N_f \ . 
\label{betaydef}
\eeq
Since
\beq
\beta_\alpha = 8\pi a^2 \bigg [ b_{1,1} N_f + \sum_{\ell=2}^\infty 
\check b_\ell \, y^{\ell-1} \bigg ] \ , 
\label{betamixed}
\eeq
one has 
\beq
\beta_y = 8\pi b_{1,1} \, y^2 \bigg [ 1 + \frac{1}{N_f} 
\sum_{\ell=2}^\infty \acute b_\ell \, y^{\ell-1} \bigg ] \ . 
\label{betay}
\eeq
The $n$-loop rescaled beta function, $\beta_{y,n\ell}$, is given by
Eq. (\ref{betay}) with the upper limit on the summation over loop order $\ell$
set to $\ell=n$ rather than $\ell=\infty$.  We will next use Eq. (\ref{betay})
below for a general analysis of a UV zero of $\beta_y$ in the large-$N_f$
limit.

The fact that $b_1$ scales differently in the LNF limit from $b_\ell$ for $\ell
\ge 2$, gives rise to significant differences between the large-$N_c$ limit in
an SU($N_c$) gauge theory and the large-$N_f$ limit in either a U(1) or
SU($N_c$) theory. Thus, in contrast to an equation for a vanishing $\beta$
function in the large-$N_c$ limit in QCD, which can be written completely in
terms of the variable $\xi = N_c \alpha$ (with infrared zeros that have been
analyzed to higher-loop order in \cite{lnn}), Eq. (\ref{betayzero}) below
contains explicit dependence on $N_f$.


\subsection{Criterion and Approximate Solution for UV Zero of $\beta_y$ in 
the Large-$N_f$ Limit} 

The condition that the $n$-loop $\beta_y$ function vanishes away from the
origin $y=0$ is the algebraic equation (of degree $n-1$ in $y$)
\beq
1 +  \frac{1}{N_f} \sum_{\ell=2}^n \acute b_\ell \, y^{\ell-1} = 0 \ . 
\label{betayzero}
\eeq
Consequently, in the LNF limit, of the $n-1$ roots of Eq.  (\ref{betayzero}),
the relevant one has the approximate form
\beq 
y_{_{UV,n\ell}} \sim
\bigg ( -\frac{N_f}{\acute b_{n,n-1}} \bigg )^{\frac{1}{n-1}} = 
\bigg ( -\frac{b_{1,1}N_f}{b_{n,n-1}} \bigg )^{\frac{1}{n-1}} \ . 
\label{ysol}
\eeq
In terms of the coupling $a=y/N_f$, or equivalently, $\alpha$, this approximate
solution is 
\beq
a_{_{UV,n\ell}} = \frac{\alpha_{_{UV,n\ell}}}{4\pi} \sim 
\bigg (-\frac{b_{1,1}}{b_{n,n-1}} \bigg )^{\frac{1}{n-1}}\, N_f^{-\frac{n-2}{n-1}} 
\ . 
\label{asol}
\eeq
This is proved as follows. 
For fixed $n$, in the LNF limit, taking account of the
prefactor $(N_f)^{-1}$ multiplying the sum, the $\ell$'th term in
Eq. (\ref{betayzero}), where $2 \le \ell \le n$, behaves like
\beq
(N_f)^{-1} \, \acute b_\ell \, y^{\ell-1} \sim \acute b_\ell \,
(-\acute b_n)^{-\frac{\ell-1}{n-1}} \, (N_f)^{-\frac{n-\ell}{n-1}} \ .
\label{termell}
\eeq
Hence, for fixed $n$, all of the $n-2$ terms with $2 \le \ell \le n-1$
vanish in the LNF limit, and the equation reduces to just
\beq
1 + ... + (N_f)^{-1} \, \acute b_\ell \, y^{n-1} = 0 \ ,
\label{betayzeroreduced}
\eeq
where the $...$ refer to the negligibly small terms in this limit. The solution
to Eq. (\ref{betayzeroreduced}) is the result in (\ref{ysol}), or equivalently,
(\ref{asol}).  Since $b_1$ and, for NFI scheme transformations, also
$b_{n,n-1}$, are scheme-independent, these solutions for $y_{_{UV,n\ell}}$ in
Eq. (\ref{ysol}) and for $a_{_{UV,n\ell}}$ in Eq. (\ref{asol}) are also NFI
scheme-independent. The fact that this solution for $N_f \to \infty$ is a
reasonably accurate approximation at large but finite $N_f$ can be demonstrated
as follows.  For values of $n$ and $N_f$ for which $\beta_{n\ell}$ has a UV
zero and for which $b_n$ is negative, we define
\beq
\kappa_{n\ell} \equiv a_{_{UV,n\ell}} 
\bigg (-\frac{\check b_n}{b_{1,1}} \bigg )^{\frac{1}{n-1}} \, 
N_f^{\frac{n-2}{n-1}} \ . 
\label{kappa_nloop}
\eeq
Recalling Eq. (\ref{bellleading}) and using the actual solutions to
Eq. (\ref{betayzero}), we then check to see that, for a fixed $n$,
$\kappa_{n\ell}$ is consistent with an approach to 1 in the LNF limit.  We
display illustrative numerical results extending from $N_f=1$ to $N_f=10^4$ in
Table \ref{kappavalues}, which show that, indeed, this is the case.  Note that
at the five-loop level, even for the interval of values $5 \le N_f \le 19$
where Eq. (\ref{betayzero}) has a solution for a UV zero, no physical 
$\kappa_{5\ell}$ is defined because $b_5$ is positive. 

From Eq. (\ref{ysol}) or (\ref{asol}),
we derive the following criterion for the existence of a UV zero in the beta
function of a U(1) gauge theory for large $N_f$: 
\beq
{\rm As} \ N_f \to \infty, \ \beta_{\alpha,n\ell} \ {\rm has \ a \ 
UV \ zero \ if \ and \ only \ if } \ b_{n,n-1} < 0 \ .
\label{criterion}
\eeq
and similarly with $\beta_y$.  Since the analysis leading to this result
extends to the non-Abelian gauge theory also, with obvious changes (replacement
of $N_f$ by $T_fN_f$ and $y=N_fa$ by $\eta=T_fN_fa$; see Eq. (\ref{lnf_nagt})),
the criterion (\ref{criterion}) also applies to a non-Abelian gauge theory. We
have shown above that $b_{n,n-1}$ is NFI scheme-independent, and hence, so is
this criterion.

The asymptotic solution (\ref{ysol}) for $y_{_{UV,n\ell}}$ has some important
implications for relating calculations of a UV zero in the $n$-loop beta
function, $\beta_{\alpha,n\ell}$ at large $N_f$ with results obtained by
summing over Feynman diagrams up to infinitely many loops in the
LNF limit (\ref{lnf}).  Let us assume that for some $n$, $b_{n,n-1} < 0$, so
that $\beta_{\alpha,n\ell}$ does have a UV zero for large $N_f$.  From
Eq. (\ref{ysol}), it follows that, for fixed loop order $n$ and 
$N_f \to \infty$, 
\beq
\lim_{N_f \to \infty} y_{_{UV,n\ell}} = \infty \ .
\label{ylnflimit}
\eeq
This property prevents one from simply matching the solution for
$y_{_{UV,n\ell}}$ (for a given $n$ for which $b_{n,n-1} < 0$ so that it exists)
to a possible solution for a finite $y$ in the LNF limit obtained by summing an
infinite set of diagrams.


\subsection{Relations Between Coefficients in Large-$N_f$ Expansion of 
$\beta_y$ and Coefficients in Small-$\alpha$ Expansion of $\beta_\alpha$} 

Expanding each $b_\ell$ in Eq. (\ref{beta}) in terms of its individual terms
multiplying different powers of $N_f$, Eq. (\ref{bellexpansion}), we have 
\beqs
& & \beta_\alpha = 8 \pi a^2 \bigg [ b_1 + 
\sum_{\ell=2}^\infty a^{\ell-1} b_\ell \bigg ] \cr\cr
& = & 8 \pi a^2 \bigg [ b_{1,1} N_f + \sum_{\ell=2}^\infty y^{\ell-1} 
( N_f)^{1-\ell} \sum_{k=1}^{\ell-1} b_{\ell,k} N_f^k \Big ] \cr\cr
      & = & 8 \pi a^2 N_f b_{1,1} \bigg [ 1 + 
 \sum_{\ell=2}^\infty y^{\ell-1} N_f^{-\ell}
\sum_{k=1}^{\ell-1} \acute b_{\ell,k} N_f^k \bigg ] \ , 
\cr\cr
& & 
\eeqs
so that 
\beq
\beta_y = 8\pi b_{1,1} \, y^2 \bigg [ 1 +
\sum_{\ell=2}^\infty y^{\ell-1} \sum_{k=1}^{\ell-1} 
\acute b_{\ell,k} N_f^{-(\ell-k)} \bigg ] \ . 
\label{betaybellk}
\eeq

An alternate way to express $\beta_y$ in the LNF limit is as an expansion
in $1/N_f$ around $1/N_f=0$.  Since one is interested here in the large-$N_f$
limit, it will be convenient to define the variable 
\beq
\nu \equiv \frac{1}{N_f} \ . 
\label{nu}
\eeq
We thus write
\beq
\beta_y = 8 \pi b_{1,1} \, y^2 \bigg [ 1 + \sum_{s=1}^\infty \frac{F_s(y)}
{N_f^s} \bigg ] = 8 \pi b_{1,1} \, y^2 \Big [ 1+\sum_{s=1}^\infty F_s(y) \nu^s
\Big ] \ . 
\label{betay_nfseries}
\eeq
By relating this expansion to the expansion Eq. (\ref{betaybellk}), we find
that if one expands $F_s(y)$ as a series in powers of $y$ about $y=0$, the term
of lowest degree in $y$ has degree $s$.  We thus write 
\beq
F_s(y) = \sum_{p=s}^\infty f_{s,p} \, y^p \ . 
\label{fsypexpansion}
\eeq

Substituting this expansion in Eq. (\ref{betay_nfseries}), we have
\beq
\beta_y = 8 \pi b_{1,1} \, y^2 \bigg [ 1 + \sum_{s=1}^\infty N_f^{-s}
  \sum_{p=s}^\infty f_{s,p} \, y^p \bigg ] \ . 
\label{betay_nfypseries}
\eeq
Matching the terms involving a given inverse power of $N_f$ in Eqs. 
(\ref{betay_nfseries}) and (\ref{betaybellk}), i.e., the term 
$f_{s,p} \, y^p N_f^{-s}$ in Eq. (\ref{betaybellk}) and the term 
$\acute b_{\ell,k} \, y^{\ell-1} N_f^{-(\ell-k)}$, we have $p=\ell-1$ and
$s=\ell-k$, and the identification
\beq
f_{s,p} = \acute b_{p+1,p+1-s} = \frac{3}{4}b_{p+1,p+1-s}
\label{fsp_to_bellk}
\eeq
or equivalently, 
\beq
b_{\ell,k} = b_{1,1} \, f_{\ell-k,\ell-1} = \frac{4}{3} f_{\ell-k,\ell-1} \ .
\label{bellk_to_fsp}
\eeq
In these equations, it is understood that the indices $\ell$, $k$, $s$, and 
$p$ range over the values in the respective expansions.  Although no 
coefficients with $s=0$ appear in Eq. (\ref{betay_nfypseries}), we 
formally define $f_{0,0} \equiv (3/4)b_{1,1} = 1$.  It will be convenient
to define
\beq
\grave f_{s,p} \equiv b_{1,1} f_{s,p} = \frac{4}{3} f_{s,p} \ . 
\label{fspgrave}
\eeq

Thus, combining these results, we have, explicitly,
\beq
F_s(y) = \sum_{s=p}^\infty f_{s,p} \, y^p = 
\frac{3}{4} \sum_{s=p}^\infty b_{p+1,p-1-s} \, y^p
\label{fsyexpansion_relto_b}
\eeq
and
\beq
b_\ell = \sum_{k=1}^{\ell-1} b_{\ell,k} \, N_f^k = 
\frac{4}{3} \sum_{k=1}^{\ell-1} f_{\ell-k,\ell-1} \, N_f^k \ .
\label{bellexpansion_relto_f}
\eeq
%


\subsection{Determination of Coefficient $b_{n,n-1}$ and Application of
Criterion for UV Zero} 

In this section we will use an exact integral representation for $F_1(y)$ from
\cite{pascual} to determine the coefficient of the term in $b_n$ of highest
degree in $N_f$, namely, $b_{n,n-1}$ and, hence to determine the existence or
non-existence of a UV zero in $\beta_\alpha$ at loop order $n$ for large $N_f$.
The integral representation for $F_1(y)$ (in the $\overline{\rm MS}$ scheme) is
\cite{pascual}
\beq
F_1(y) = \int_{0}^{\frac{4y}{3}} dx \, I_1(x) \ , 
\label{f1}
\eeq
where
\beq
I_1(x) = \frac{(1+x)(1-2x)(1-2x/3)\Gamma(4-2x)}{b_{1,1} \Gamma(1+x)\Gamma(3-x)
[\Gamma(2-x)]^2} \ , 
\label{i1}
\eeq
Here $\Gamma(x)$ is the Euler gamma function. Some implications of this for a
possible nonperturbative zero in the U(1) $\beta$ function have been discussed
in \cite{holdom2010}.

From an inspection of Eq. (\ref{i1}), one sees that the integrand $I_1(x)$ is
analytic in the complex $x$ plane in a disk around $x=0$ of radius $|x|=5/2$.
One can therefore expand this integrand function in a Taylor series about $x=0$
and integrate term by term.  The resultant integral, which is the function
$F_1(y)$, is analytic in the complex $y$ plane in a disk around $y=0$ of radius
$|y|=15/8$.  We thus obtain a Taylor series for $F_1(y)$ around $y=0$ to
arbitrarily high order. Now we combine this with the relation
(\ref{fsp_to_bellk}) derived above equating $f_{s,p}$ with $\acute
b_{p+1,p+1-s}$.  We take the $s=1$ special case of this relation to get
\beqs
f_{1,p} & = & \acute b_{p+1,p} = \frac{3}{4} b_{p+1,p} \ , \ i.e., 
\cr\cr
b_{\ell,\ell-1} & = & \grave f_{1,\ell-1} = \frac{4}{3} \, f_{1,\ell-1}
\ . 
\label{fsp_to_bellk_highestNf}
\eeqs

Combining this equation expressing $f_{1,p}=(3/4)b_{p+1,p}$ with the result
proved above, that the $b_{\ell,\ell-1}$ are invariant under NFI scheme
transformations, one proves that the $f_{1,p}$ are invariant under these scheme
transformations.  Since these coefficients determine $F_1(y)$ via the $s=1$
special case of the Taylor series expansion (\ref{fsypexpansion}), it follows
that $F_1(y)$ is NFI scheme-independent.  Although the $F_1(y)$ function is
different for a non-Abelian gauge theory (in the latter case, we denote it as
$F_1^{(NA)}(\eta)$, where $\eta$ is defined below in Eq. (\ref{lnf_nagt})), the
steps in the proof of NFI scheme-invariance are the same, so the result holds
for both an Abelian and non-Abelian gauge theory. Thus,
\beq
F_1(y)_{ST} = F_1(y), \quad F_1^{(NA)}(\eta)_{ST} = F_1^{(NA)}(\eta) \ , 
\label{f1invariance}
\eeq
where the subscript $ST$ means the function after the NFI scheme transformation
in Eqs. (\ref{aap}) and (\ref{faprime}) has been applied.  

Since the Taylor series expansion of $F_1(y)$ yields all $f_{1,p}$ for
arbitrarily large $p$, these determine the $b_{p+1,p}$, i.e.,
$b_{\ell,\ell-1}$, namely the terms in $b_\ell$ of highest degree in $N_f$ for
arbitrarily large loop order $\ell$.  We list some of the low-order
coefficients $f_{1,p}$ in Appendix \ref{f1taylor} for reference. 
It is readily verified that the
results for $b_{\ell,\ell-1}$ obtained via this Taylor series expansion in
conjunction with Eq. (\ref{fsp_to_bellk_highestNf}) agree with the known
results from $b_\ell$ for loop orders $\ell=2$ to $\ell=5$, namely,
$b_{2,1}= \grave f_{1,1}=4$, as in Eq. (\ref{b21}) above, 
\beq
b_{3,2}= \grave f_{1,2}=-\frac{44}{9}=-4.88889 \ ,
\label{b32}
\eeq
\beq
b_{4,3} = \grave f_{1,3} = -\frac{1232}{243}=-5.06996 \ , 
\label{b43}
\eeq
and
\beq
b_{5,4} = \grave f_{1,4} = \frac{856}{243}+\frac{128\zeta(3)}{27} = 
9.22127 \ , 
\label{b54} 
\eeq
where here and below, we leave factorizations of integers implicit, and
numerical values are given to the indicated floating-point accuracy.  

Going beyond these known coefficients, we have calculated analytic expressions
for the $b_{\ell,\ell-1}$ to higher loop order $\ell$.  For $\ell = 6, \ 7, \
8$ we obtain
\beq
b_{6,5} = \grave f_{1,5} =  \frac{16064}{3645} - \frac{11264\zeta(3)}{1215}
+ \frac{512\pi^4}{6075} = 1.47275
\label{b65}
\eeq
\beqs
b_{7,6} & = & \grave f_{1,6} = \frac{4288}{729} - \frac{78848\zeta(3)}{6561}
- \frac{5632\pi^4}{32805} + \frac{4096\zeta(5)}{243} \cr\cr
& = & -7.80879
\label{b76}
\eeqs
and
\beqs
b_{8,7} &=& \grave f_{1,7} = \frac{52480}{6561} + \frac{438272\zeta(3)}{45927}
- \frac{22528\pi^4}{98415} \cr\cr
& - & \frac{180224\zeta(5)}{5103} + \frac{32768\pi^6}{964467}
+ \frac{32768\zeta(3)^2}{5103}  \cr\cr
& = & 2.49243 \ . 
\label{b87}
\eeqs
In Appendix \ref{bnnminus1} we list the analytic expressions that we have
calculated for the next two $b_{\ell,\ell-1}$ coefficients, namely those for
loop order $\ell=9$ and $\ell=10$.  These analytic expressions for the
$b_{\ell,\ell-1}$ with $6 \le \ell \le 10$ are new results here.  As is
evident, the expressions become rather lengthy as $\ell$ increases.  The
numerical values of $b_{\ell,\ell-1}$ are given for loop order $\ell$ up to 24
in Table \ref{bellnfmax} (with $\ell \equiv n$ in the notation).  This extends
the numerical values given in \cite{pascual}. (Note that the coefficient
$(\beta)_n$ in \cite{pascual} is equal to $4^{-(n+1)}b_{n+2,n+1}$ in our
notation.)

From Eq. (\ref{fsp_to_bellk_highestNf}) and Eq. (\ref{criterion}), it follows
that the sign of $b_{n,n-1}$ determines whether, for large $N_f$, there is a UV
zero in the $n$-loop beta function, $\beta_{\alpha,n\ell}$; this UV zero exists
if and only if ${\rm sgn}(b_{n,n-1})$ is negative. Thus, the results listed in
Table \ref{bellnfmax} determine whether the $n$-loop beta function has a UV
zero for large $N_f$.  For example, since $b_{3,2}$ is negative, the three-loop
beta function, $\beta_{\alpha,3\ell}$ has a UV zero for large $N_f$, and the
same is true for the four-loop beta function, $\beta_{\alpha,4\ell}$, since
$b_{4,3}$ is also negative.  But in contrast, since $b_{5,4}$ and $b_{6,5}$ are
positive, the respective five-loop and six-loop beta functions do not have any
UV zero at large $N_f$, etc. In order for the $\beta_\alpha$ function to show
evidence for a (stable) UV zero at large $N_f$, $\beta_{\alpha,n\ell}$ should
exhibit such a zero, i.e., the signs of the $b_{n,n-1}$ coefficients should be
negative.  Instead, as is evident in Table \ref{bellnfmax}, one finds a
scattering of both positive and negative values of $b_{n,n-1}$ up to $n=24$.
Thus, up to the $n=24$ loop level that we have calculated, the $n$-loop $\beta$
function of the U(1) theory does not exhibit a stable UV zero for large $N_f$.
There is the possibility that as the loop order $n$ increases beyond some value
greater than 24, the $b_{n,n-1}$ will become uniformly negative, so that 
the corresponding $n$-loop beta function, $\beta_{\alpha,n\ell}$ will develop
at stable UV zero at large $N_f$. However, one would still have to contend with
the result (\ref{ylnflimit}), which does not yield a finite value of
$y_{_{UV,n\ell}}$ as $N_f \to \infty$ for fixed $n$.  These findings do not
give support for a UV zero of the beta function in the LNF limit. However, they
do not rigorously preclude such a UV zero obtained by the calculation of a
leading infinite set of fermion vacuum polarization insertions on gauge boson
propagator lines.  If such a UV zero should exist, it would mean that
calculations of the $n$-loop beta function for fixed finite $n$ and large $N_f$
are not sensitive to it.  Some insight into this possibility is gained from a
comparison of the $b_{n,n-1}$ coefficients with certain coefficients 
$b_{n,n-1}^{(d)}$ defined in Eq. (\ref{bnnminus1fdiv}) below. 


\subsection{Calculations of Higher Coefficients $f_{s,p}$ for $2 \le s \le 4$} 

Although the $F_s(y)$ with $s \ge 2$ have not been calculated exactly, we can
determine some of the coefficients $f_{s,p}$ in these $F_s(y)$ by reversing the
procedure used in the previous section, using Eq. (\ref{fsp_to_bellk}) together
with the known coefficients $b_\ell$. For general $s$, Eq. (\ref{fsp_to_bellk})
gives $f_{s,p}$ in terms of $\acute b_{p+1,p+1-s}$, so if the $b_\ell$ have
been calculated up to $\ell$-loop order, then this relation enables one to
determine the $f_{s,p}$ for $s \le p \le \ell-1$.  Since the $b_\ell$ have been
computed up to $\ell=n=5$ loops (in the $\overline{\rm MS}$ scheme)
\cite{b5u1a,b5u1b}, one can thus determine $f_{s,p}$ for $s \le p \le 4$ for
$F_s(y)$ with $2 \le s \le 4$. 

We consider $F_2(y)$ first.  Here, Eq. (\ref{fsp_to_bellk}) gives
$f_{2,p}=\acute b_{p+1,p-1}$ with $p \ge 2$. The first two of coefficients 
$f_{2,p}$ are known: 
\beq
f_{2,2} = \acute b_{3,1} = -\frac{3}{2}  \ , 
\label{f22}
\eeq
and
\beq
f_{2,3} = \acute b_{4,2} = \frac{190}{9} - \frac{208\zeta(3)}{3} 
= -62.23150 \ . 
\label{f23}
\eeq
For $f_{2,4}$, we find \cite{holdif}:
\beqs
f_{2,4} & = & \acute b_{5,3} = \frac{-10879}{54} + \frac{4000\zeta(3)}{9}
-\frac{52\pi^4}{45} - 320\zeta(5) \cr\cr
& = & -(1.1159395 \times 10^2) \ . 
\label{f24}
\eeqs

We next study $F_3(y)$.  Setting $s=3$ in Eq. (\ref{fsp_to_bellk}) gives
$f_{3,p}=\acute b_{p+1,p-2}$ with $p \ge 3$. Here,
\beq
f_{3,3} = \acute b_{4,1} = -\frac{69}{2} = -34.5 
\label{f33}
\eeq
and
\beqs
f_{3,4} & = & \acute b_{5,2} = -\frac{3731}{6}-744\zeta(3)+2040\zeta(5)
\cr\cr
        & = & 5.9916895 \times 10^2 \ . 
\label{f34}
\eeqs

Finally, we consider $F_4(y)$. Setting $s=4$ in Eq. (\ref{fsp_to_bellk}) yields
$f_{4,p}=\acute b_{p+1,p-3}$ with $p \ge 4$. Thus, 
\beqs
f_{4,4} & = & \acute b_{5,1} = \frac{4157}{8}+96\zeta(3) \cr\cr
        & = & 6.3502246 \times 10^2 \ . 
\label{f44}
\eeqs
%


\subsection{Role of $F_1(y)$ Regarding a Possible UV Zero of the Beta 
Function } 

We address here the question of whether $\beta_y$ might have a UV zero for
large $N_f$.  From Eq. (\ref{betay_nfseries}), one can make several general
statements.  First, since the LNF limit of Eq. (\ref{lnf}) is $N_f \to \infty$
with $y$ held fixed and finite, if the value of $y$ is such that the $F_s(y)$
are finite, then each terms, $F_s(y)/N_f^s$ in the series $\sum_{s=1}^\infty$
vanishes, and $\beta_y$ reduces to just the first term, $\beta_y=8\pi y^2
b_{1,1} = (32\pi/3)y^2$, which does not vanish for any nonzero $y$.  The
question then is whether for some large but finite $N_f >> 1$, $\beta_y$ might
vanish.  

As is evident from Eq. (\ref{betay_nfseries}), to order $O(1/N_f)$, the
condition that $\beta_y$ vanishes is that $1+F_1(y)/N_f=0$.  This condition is
satisfied for large $N_f$ if and only if there is a value of $y$ for which
$F_1(y)$ is very large and negative.  Now the expansion variable $1/N_f$ in
Eq. (\ref{betay_nfseries}) (for fixed $y$) can be regarded as being formally
analogous to the expansion variable $a$ or $\alpha$ in Eq. (\ref{beta}) (for
fixed $N_f$).  

One may ask whether the necessary (and sufficient) condition that, to order
$1/N_f$, $\beta_y$ can vanish, i.e., that there exists a value of $y$ such that
$F_1(y)$ is large and negative, is satisfied.  Using the exact result from
\cite{pascual}, Holdom observed that this is, indeed, the case
\cite{holdom2010}. We recall his analysis. The integrand function $I_1(x)$ in
Eq. (\ref{i1}) has simple poles at
\beq
x_{pole} = \frac{5}{2} + p \quad {\rm for} \ p = 0, \ 1, \ 2, \ 3,...
\label{xpole}
\eeq
These produce divergences in $F_1(y)$.  The first of these is a logarithmic
divergence at
\beq
y_{div} = \frac{15}{8} \ . 
\label{ylogdiv}
\eeq
Next, one expands the integrand around this singularity, uses the
Taylor-Laurent series for $\Gamma(x)$ for $x=-n+\epsilon$, with $n=0, \ 1, \ 2,
...$ and $\epsilon \to 0$ (see Appendix \ref{gammafunction}), and integrates
the result term by term. One finds that for $y$ slightly less than $y_{div}$,
the logarithmically divergent part of $F_1(y)$ is
\beq
F_1(y)_{div} \sim \frac{7}{15\pi^2} \, \ln | 1 - (y/y_{div})| + {\rm finite} 
\ . 
\label{f1logdiv}
\eeq
Since this diverges negatively as $y \nearrow y_{div}$, it follows that there
exists a value of $y$, which we denote $y_0$, for which the term $F_1(y)/N_f$
is equal to $-1$, so that $\beta_y=0$ to this order in the $1/N_f$ expansion.
This value $y_0$ is exponentially close to the value $y_{div}$ where $F_1(y)$
has a (logarithmic) divergence;
\beq
y_0 \simeq \frac{15}{8} - {\rm const.} \times e^{-15\pi^2N_f/7}  \ . 
\label{ybetayzero}
\eeq

Obviously, one must treat this result with caution, because of the divergence
in $F_1(y)$ at $y_{div}$. However, it suggests that for large $N_f$, $\beta_y$
might have a UV zero.  Somewhat analogously to the situation with the nonlinear
$\sigma$ model \cite{nlsm}, the existence of this UV zero would be the result
of having summed an infinite subset of Feynman diagrams that are dominant in
the limit $N_f \to \infty$.  One way to study a possible zero in $\beta_y$
further entails an analysis of contributions to $\beta_y$ from terms that are
of higher order in $1/N_f$, namely, $\sum_{s=2}^\infty F_s(y)/N_f^s$. Since the
$F_s(y)$ for $s \ge 2$ have not been calculated exactly, this analysis is,
perforce, exploratory. The $F_s(y)$ with $s \ge 2$ are also NFI
scheme-dependent, in contrast with $F_1(y)$. It has been suggested that in the
interval $I_y=[0,y_{div}]$, $F_2(y)$ might contain a pole at $y=y_{div}$, the
$F_s(y)$ with $s \ge 2$ might have higher-order poles, and the effect of these
might be to remove the UV zero in $\beta_y$ in this interval obtained with just
the leading $F_1(y)/N_f$ term \cite{holdom2010}.  That is, for a fixed large
value of $N_f$, as $y \to y_{div}$, higher-order terms in the expansion
(\ref{betay_nfseries}) could dominate over the $F_1(y)/N_f$ term. 
For fixed $y$ slightly different from $y_{div}$, provided that the
$F_s(y)$ with $s \ge 2$ are finite at this value of $y$, one can make $N_f$
large enough to reduce the contributions of any finite set of terms
$F_s(y)/N_f^s$ with $s \ge 2$ to values smaller than $F_1(y)/N_f$.  But this
does not necessarily hold for the infinite series $\sum_{s=2}^\infty
F_s(y)/N_f^s$.  In our analysis above, we have used a complementary method,
namely to investigate the presence or absence of a UV zero in
$\beta_{\alpha,n\ell}$ at large $N_f$.

We note that a possible UV zero in $\beta_y$ would be different from, for
example, an IR zero in the two-loop beta function of a non-Abelian (NA) gauge
theory, $\beta_{\alpha,2\ell}^{(NA)} =
(\alpha^2/(2\pi))(b_1^{(NA)}+b_2^{(NA)}a)$ at $a_{_{IR,2\ell}} =
-b_1^{(NA)}/b_2^{(NA)}$, where $b_1^{(NA)} < 0$ and $b_2^{(NA)} > 0$, because,
while the latter occurs at a fixed value of $a$, here the zero would occur at a
fixed value of $y=N_fa$, with $a$ itself having been driven toward zero because
of the condition that $y$ be fixed as $N_f \to \infty$ in the LNF limit.
A second difference can be seen as follows. By either analytically continuing
$N_f$ from non-negative integer values to non-negative real values or by
keeping $N_f$ fixed at positive integer values and letting $N_c$ become large,
one can choose $N_f$ to be very slightly less than the value at which
$b_1^{(NA)}$ goes through zero and reverses sign, and hence one can make the
value of an IR zero of $\beta_\alpha^{(NA)}$ arbitrarily small.  This is not
the case with a possible UV zero of $\beta_y$ calculated to a given order in
$1/N_f$. A third difference, closely related to the second, is that,
whereas the condition for an IR zero in $\beta_\alpha^{(NA)}$, say at two
loops, viz., $b_1^{(NA)}+b_2^{(NA)} a=0$, does not require $b_2^{(NA)}$ to be
extremely large, since $b_1^{(NA)}$ can be quite small, the condition for a
zero in $\beta_y$ at large $N_f$ does generically require $|F_1(y)|$ to be very
large.


\subsection{Calculations of the Coefficients $b_{n,n-1}^{(d)}$ and Comparison
  with the $b_{n,n-1}$} 

Since the large negative value of $F_1(y)$ as $y$ approaches $y_{div}$ from
below is a key feature in the possibility of a zero in $\beta_y$ to order
$1/N_f$, it is interesting to investigate how close the actual calculated
values of $f_{1,p}$ and hence, by Eq. (\ref{bellk_to_fsp}), the $b_{n,n-1}$
with $n=p+1$ for the U(1) gauge theory are to the values that one would get by
calculating a Taylor series expansion (around $y=0$) of the leading divergent
term in Eq. (\ref{f1logdiv}).  The coefficients in this Taylor series expansion
are
\beq
f_{1,p}^{(d)} = \frac{1}{p!} \frac{d^pF_1(y)_{div}}{dy^p}{}\bigg |_{y=0} 
\label{f1pdiv}
\eeq
(where $d$ stands for $div$), so, in analogy with Eq. (\ref{fsypexpansion}), 
\beq
F_1(y)_{div} = \sum_{p=1}^\infty f_{1,p}^{(d)} \, y^p \ . 
\label{f1ydivseries}
\eeq
From Eq. (\ref{fsp_to_bellk_highestNf}), one then has the corresponding
coefficients  
\beq
b_{n,n-1}^{(d)} = \frac{4}{3} f_{1,n-1}^{(d)} \ . 
\label{bnnminus1fdiv}
\eeq
We list the $b_{n,n-1}^{(d)}$ in Table \ref{bnminus1div} for $n$ up to 24.  The
$f_{1,p}^{(d)}$, and hence $b_{n,n-1}^{(d)}$ with $n=p+1$, are all negative,
and they decrease in magnitude monotonically and sufficiently with increasing
$n$ as to be consistent with a finite limit above unity for large, fixed $N_f$:
\beq
\lim_{n \to \infty} 
\bigg ( - \frac{4N_f}{3b_{n,n-1}^{(d)}} \bigg   )^{\frac{1}{n-1}} = 
\lim_{n \to \infty} 
\bigg ( - \frac{N_f}{f_{1,n-1}^{(d)}}    \bigg  )^{\frac{1}{n-1}} \ . 
\label{ysoldivlim}
\eeq

An important conclusion follows from comparing the results for the series
expansion coefficients $b_{n,n-1}$ for the full function $F_1(y)$, listed in
Table \ref{bellnfmax}, with the corresponding coefficients $b_{n,n-1}^{(d)}$
that we have calculated from $F_1(y)_{div}$ and listed in Table
\ref{bnminus1div}.  Two key differences are evident, namely that (i) while
the $b_{n,n-1}$ exhibit a scattering of positive and negative signs, the
$b_{n,n-1}^{(d)}$ are uniformly negative, and (ii) while the $|b_{n,n-1}|$ do
not decrease monotonically, the $|b_{n,n-1}^{(d)}|$ do, as discussed above.  We
conclude from these differences that the series expansion in $y$ for the full
function $F_1(y)$, up to order $n=24$, is not dominated by the effect of
the logarithmic divergence at $y=y_{div}=15/8$. 


\section{Implications from Illustrative Test Functions for $\beta_y$} 
\label{testfunctions} 

In this section we consider several illustrative test functions for $\beta_y$
that, by construction, for fixed $\nu=1/N_f$, have a UV zero at a value of $y$
that we denote as $y_z$ (where the subscript stands for ``zero'').  We explore
how the properties of these test functions are manifested in the two types of
series expansions studied here, namely the expansion in powers of $\nu$ around
the point $\nu=0$ for fixed $y=N_fa$ in Eq.  (\ref{betay_nfseries}) and the
expansion in powers of $a$, or equivalently $\alpha$, around $a=0$ for fixed
large $N_f$ in Eq. (\ref{beta}).  The choice of these test functions is
restricted by some general properties.  We recall from our discussion above
that the term of lowest degree in the expansion of $F_s(y)$ about $y=0$ has
degree $s$ in $y$.  The factor $[1+\sum_{s=1}^\infty F_s(y) \nu^s]$ in
Eq. (\ref{betay_nfseries}) has the properties that (i) it is equal to 1 for
arbitrary $y$ if $\nu=0$, and (ii) it is equal to 1 for arbitrary $\nu$ if
$y=0$.  We define a function $\Phi(y,\nu)$ as
\beq
\Phi(y,\nu) \equiv \frac{\beta_y}{4\pi b_{1,1} y^2} = \frac{3\beta_y}{16 \pi
  y^2} \ . 
\label{Phi}
\eeq
The function $\Phi(y,\nu)$ thus satisfies the properties 
\beq
\Phi(y,0)=1
\label{Phiy0}
\eeq
and
\beq
\Phi(0,\nu)=1 \ . 
\label{Phi0nu}
\eeq
As noted above, since our purpose in constructing and analyzing these
$\Phi(y,\nu)$ test functions is to investigate their implications for a
possible UV zero in the actual beta function of a U(1) gauge theory, they are
designed to have a zero at $y=y_z$, as approached from below (in an interval
connected with small $y$), i.e.,
\beq
\lim_{y \nearrow y_z} \Phi(y,\nu) = 0 \ .
\label{Phizero}
\eeq
We will study these test functions for $y$ in the interval 
\beq
I_y: \quad 0 \le y \le y_z 
\label{yinterval}
\eeq
and focus on the region near $\nu=0$, i.e., large $N_f$.  The functions 
$F_s(y)$ in Eq. (\ref{betay_nfseries}) are then given by 
\beq
F_s(y) = \frac{1}{s!} \frac{d^s \Phi(y,\nu)}{d\nu^s}{}\Bigg |_{\nu=0} \ .
\label{fsPhi}
\eeq
Note that in taking this derivative, it is necessary to extend the definition
of $\nu$ from (a subset of) ${\mathbb Q}_+$ to ${\mathbb R}$; this extension is
to be implicitly understood where necessary.  Our analysis here also applies to
the non-Abelian case to be discussed in the next section, with the replacement
of $y$ by $\eta$ as given in Eq. (\ref{lnf_nagt}).  The question of what
analytic form $\Phi(y,\nu)$ takes for $y > y_z$ is beyond the scope of our
study here, since we are only interested in the behavior in the interval $I_y$,
i.e., the question of a UV zero that would be reached by the coupling via
renormalization-group evolution, starting from an initial small value in the
infrared.  Although, by design, our $\Phi(y,\nu)$ test functions satisfy the
necessary conditions (\ref{Phiy0})-(\ref{Phizero}), and it is hoped that they
give insight into the behavior of the true function $\Phi(y,\nu)$, no
implication is made that they are fully realistic.

A particularly simple test function, applicable in the interval $I_y$, 
is the power-law form
\beq
\Phi(y,\nu) = [1-(y/y_z)]^\nu \ , 
\label{Phi_powerlawzero}
\eeq
where $\nu$ is a positive number. With this test function, we now calculate the
resultant functions $F_s(y)$ appearing in the expansion (\ref{betay_nfseries}).
Since 
\beq
[1-(y/y_z)]^\nu = e^{\nu \ln[1-(y/y_z)]} = 1+\sum_{s=1}^\infty 
\frac{\nu^s}{s!}\Big [ \ln \Big ( 1-(y/y_z) \Big ) \Big ]^s \ , 
\label{Phi_powerlawexp}
\eeq
it follows that 
\beq
F_s(y) = \frac{1}{s!}\Big [ \ln \Big ( 1-(y/y_z) \Big ) \Big ]^s \ . 
\label{Fs_powerlawzero}
\eeq
If we were to truncate this series with the $s=1$ term, this would be
analogous in form to the dominant negative logarithmic divergence in the actual
$F_1(y)$ for the U(1) gauge theory, as given in Eq. (\ref{f1logdiv}).  
Since $\nu$ is not, in general, an integer, and, indeed, we are interested in
the regime where $\nu$ is approaching zero, $\Phi(y,\nu)$ generically has a
branch point singularity at $y=y_z$. Thus, generically, $F_s(y)$ is analytic
about the point $y=0$ in the complex $y$ plane inside a disk of radius
$|y|=y_z$.  (If one were to set $\nu$ equal to an integer, $F_s(y)$ would be
analytic everywhere in the $y$ plane.)  We calculate the Taylor series
expansion of $F_s(y)$ around $y=0$ to get the coefficients $f_{s,p}$ in
Eq. (\ref{fsypexpansion}).  The results obey the requisite condition for the
U(1) (and non-Abelian) gauge theory that the Taylor expansion of $F_s(y)$
around $y=0$ has, as its lowest-degree term in $y$, a term proportional to
$y^s$.  For the case $s=1$ that yields the leading terms in the large-$N_f$
(small-$\nu$) limit, we calculate the coefficients $f_{1,p}$ in the expansion
of $F_1(y)$ to be
\beq
f_{1,p} = -\frac{1}{p} \ . 
\label{f1_ppowerlawzero}
\eeq

We have also constructed and studied a family of illustrative $\Phi(y,\nu)$
functions, applicable in the interval $I_y$, each of which has an essential
zero at $y=y_z$, namely,
\beq
\Phi(y,\nu)= \exp \bigg [ \nu \bigg ( -\frac{1}{(1-(y/y_z))^k} + 1 \bigg ) 
\bigg ] \ . 
\label{Phi_essentialzero}
\eeq
Here, in addition to $y_z$, $\Phi(y,\nu)$ depends on a second parameter, the
positive real number $k$, which determines the nature of the essential zero. In
particular, if $k$ is a positive integer, then $\ln[\Phi(y,\nu)]$ has a pole of
order $k$ at $y_z$.  For the $\Phi(y,\nu)$ in Eq. (\ref{Phi_essentialzero}), we
calculate
\beq
F_s(y) = \frac{1}{s!} \Bigg [ -\frac{1}{\Big (1-(y/y_z) \Big )^k} + 1 
\Bigg ]^s \ . 
\label{Fs_essentialzero}
\eeq
From this we compute the $f_{s,p}$ coefficients in Eq. (\ref{fsypexpansion}).
The results again obey, as they must, the requirement that 
$F_s(y)$ has, as its lowest-degree term in $y$, a term
proportional to $y^s$. For $s=1$, we calculate 
\beq
f_{1,p} = - {k+p-1 \choose p} \, y_z^{-p}  .
\label{f1p_essentialk}
\eeq
where ${m \choose n}=m!/[n!(m-n)!]$ is the binomial coefficient.  Although we
are only concerned here with the behavior in the interval $I_y$, we note
parenthetically that one could recast the test function
(\ref{Phi_powerlawzero}) so as to remain real for $y > y_z$ by using
$\Phi(y,\nu) = [(1-(y/y_z)^2)^2]^\nu$, and one could restrict $k$ to be even in
(\ref{Phi_essentialzero}) so that the resultant $\Phi(y,\nu)$ vanishes instead
of diverging as $y$ approaches $y_z$ from above. With these definitions,
$\Phi(y,\nu)$ would be a bounded (real, positive) function for $y > y_z$, so
that $y_z$ would be an infrared zero of $\beta_y$ for $y > y_z$. Test functions
for which $y_z$ would be an ultraviolet zero of $\beta_y$ for $y > y_z$ can
also be envisaged.  

An important property of these coefficients $f_{1,p}$ in Eq.
(\ref{f1_ppowerlawzero}) for the $\Phi(y,\nu)$ function
(\ref{Phi_powerlawzero}) and in Eq. (\ref{f1p_essentialk}) for the
$\Phi(y,\nu)$ function (\ref{Phi_essentialzero}), is that they are all
negative. From Eq. (\ref{fsp_to_bellk}) or (\ref{bellk_to_fsp}), it follows
that the $b_{n,n-1}$ for $n \ge 2$ corresponding to these $\Phi(y,\nu)$
functions are all negative.  Recalling Eq. (\ref{ysol}) and the condition
(\ref{criterion}), this property of negative $f_{1,p}$ and hence negative
$b_{n,n-1}$ coefficients resulting from both of the illustrative $\Phi(y,\nu)$
functions above implies that, regarding the other type of expansion, namely the
expansion of $\beta_\alpha$ in powers of $a$, the resultant $n$-loop beta
function $\beta_{\alpha,n\ell}$ would always exhibit a UV zero for small $\nu$
(i.e., large $N_f$).  As we have discussed and as is evident from Table
\ref{bellnfmax}, the actual $b_{n,n-1}$ coefficients for the U(1) gauge theory
are not uniformly negative, but instead exhibit a scatter of signs up to the
maximal loop order $n=24$ which we have studied.  As was noted above, this
means that, up to this loop order, these $b_{n,n-1}$ do not exhibit evidence
for a UV zero in the respective beta functions at large $N_f$.

The coefficients $f_{s,p}$ in the small-$y$ expansions of the $F_s(y)$ 
can also be calculated for higher values of $s$.
For example, for the illustrative $\Phi(y,\nu)$ function in
Eq. (\ref{Phi_powerlawzero}) with a power-law zero, we find 
\beq
F_2(y) = \frac{1}{2}y^2 + \frac{1}{2}y^3 + \frac{11}{24}y^4 + \frac{5}{12}y^5 
+ O(y^6) \ , 
\label{F2_powerlawzero}
\eeq
\beq
F_3(y) = -\frac{1}{6}y^3 - \frac{1}{4}y^4 - \frac{7}{24}y^5 - O(y^6) \ , 
\label{F3_powerlawzero}
\eeq
\beq
F_4(y) = \frac{1}{24}y^4+ \frac{1}{12}y^5+O(y^6) \ , 
\label{F4_powerlawzero} 
\eeq
and so forth for higher $s$.  For the $\Phi(y,\nu)$ with $k=1$ in
Eq. (\ref{Phi_essentialzero}), we compute
\beq
F_2(y) = \frac{1}{2}y^2 + y^3 + \frac{3}{2}y^4 + 2y^5 + O(y^6) \ , 
\label{F2_essentialzero_k1}
\eeq
\beq
F_3(y) = -\frac{1}{6}y^3 -\frac{1}{2} y^4 - y^5 - O(y^6) \ , 
\label{F3_essentialzero_k1}
\eeq
\beq
F_4(y) = \frac{1}{24}y^4 + \frac{1}{6}y^5 + O(y^6) \ , 
\label{F4_essentialzero_k1}
\eeq
and so forth for higher $s$.  We have also calculated these Taylor series
expansions of $F_s(y)$ for other values of $k$.  A general property that we
find is that for both the $\Phi(y,\nu)$ functions with a power-law zero and an
essential zero, the nonzero coefficients $f_{s,p}$ are negative for $s$ odd and
positive for even $s$, i.e., 
\beq
{\rm sgn}(f_{s,p}) = (-1)^s \ .
\label{sgnfsp}
\eeq
for these functions.  

In passing, we add a comment concerning the power-law and essential zeros in
the test function $\Phi(y,\nu)$ and hence $\beta_y$.  Although an essential
zero in $\Phi(y,\nu)$ at a given point, here, $y=y_z$, cannot be detected to
any order of a series expansion about $y_z$, since all of the coefficients
vanish identically, the zero does manifest itself in the series expansion of
$F_s(y)$ about $y=0$.  Indeed, as we have shown, the coefficients $f_{s,p}$
obtained from the expansion of $F_s(y)$ about $y=0$ share important properties
in common, such as (\ref{sgnfsp}) for both the power-law-zero form
(\ref{Phi_powerlawzero}) and the essential-zero form (\ref{Phi_essentialzero})
of $\Phi(y,\nu)$.


\section{Non-Abelian Gauge Theory}
\label{nagt}


\subsection{Structure of Beta Function} 

In this section we discuss the question of a possible UV zero of the $\beta$
function for a non-Abelian gauge theory with a sufficiently large fermion
content that this $\beta$ function is positive near the origin. We will
consider a theory with gauge group $G$ and $N_f$ massless (Dirac) fermions
transforming according to a representation $R$ of $G$.  As noted above, this
large-$N_f$ non-Abelian gauge theory is IR-free and is similar in this sense to
the U(1) gauge theory.  Let $T_f \equiv T(R)$, where $T(R)$ is the usual trace
invariant defined by \cite{casimir}
\beq
\sum_{i,j=1}^{{\rm dim}(R)} {\cal D}_R(T_a)_{ij} {\cal D}_R(T_b)_{ji} = T(R) 
\delta_{ab} \ .
\label{tr}
\eeq
where $a$ is a group index running from 1 to the order of the group, and ${\cal
D}_R(T_a)$ is the $R$-representation ($R$-{\it Darstellung}) of the generator
$T_a$ of the Lie algebra of $G$.

We consider the limit analogous to Eq. (\ref{lnf}), namely 
\beq
N_f \to \infty \quad {\rm with} \quad \eta(\mu) \equiv T_f N_f \, a(\mu) = 
\frac{T_f N_f \, \alpha(\mu)}{4\pi} \ , 
\label{lnf_nagt}
\eeq
where the function $\eta(\mu)$ is a finite function of the Euclidean scale
$\mu$ in the $N_f \to \infty$. As noted above, to distinguish the quantities
for the non-Abelian gauge theory from the analogous quantities for the Abelian
U(1) gauge theory, we will use the superscript $(NA)$.  We thus write, for the
beta function,
\beq
\beta_\alpha^{(NA)} = 2\alpha \sum_{\ell=1}^\infty b_\ell^{(NA)} \, a^\ell =
 2\alpha \sum_{\ell=1}^\infty \bar b_\ell^{(NA)} \, \alpha^\ell \ . 
\label{beta_nagt}
\eeq
For small coupling $\alpha$, since $\beta > 0$, as the
Euclidean momentum scale decreases toward the infrared, $\alpha(\mu) \to 0$,
i.e., the theory is formally free in the infrared. 

Since the coefficient of the one-loop term in $\beta^{(NA)}$ is
\cite{b1,casimir}
\beq
b_1^{(NA)} = \frac{1}{3}(-11 C_A + 4T_fN_f) \ , 
\label{b1}
\eeq
we are interested in the interval 
\beq
N_f > N_{f,b1z} \ , 
\label{nfinterval}
\eeq
where $N_{f,b1z}$ denotes the value of $N_f$ where $b_1^{(NA)}$ vanishes with
sign reversal as a function of $N_f$, namely,
\beq
N_{f,b1z} = \frac{11C_A}{4T_f} \ . 
\label{nfb1z}
\eeq

The two-loop coefficient is \cite{b2} 
\beq
b_2^{(NA)}=\frac{1}{3}\left [ - 34 C_A^2 + 4(5C_A+3C_f)T_fN_f \right ]
\ .
\label{b2}
\eeq
This is negative for small $N_f$ and vanishes with sign reversal at $N_f = 
N_{f,b2z}$, where
\beq
N_{f,b2z} = \frac{17C_A^2}{2(5C_A+3C_f)T_f} \ . 
\label{nfb2z}
\eeq
For arbitrary $G$ and $R$, $N_{f,b2z} < N_{f,b1z}$.  Hence, our restriction to
the range $N_f > N_{f,b1z}$, for which $b_1^{(NA)} > 0$, implies that
$b_2^{(NA)} > 0$ also.  Consequently, for arbitrary $G$ and $R$, this theory
has no UV zero at the two-loop level.  This is similar to the situation in the
U(1) theory.

The one-loop coefficient has the form 
\beq
b_1^{(NA)} = b_{1,0}^{(NA)} + b_{1,1}^{(NA)} \, T_fN_f \ , 
\label{b1nf_nagt}
\eeq
where
\beq
b_{1,0}^{(NA)} = -\frac{11}{3}C_A
\label{b10nagt}
\eeq
and
\beq
b_{1,1}^{(NA)}= b_{1,1} = \frac{4}{3} \ . 
\label{b11nagt}
\eeq
The coefficients $b_\ell^{(NA)}$ for $\ell \ge 2$ have the generic form of
polynomials in $(T_fN_f)$ of lowest degree 0 and highest degree $\ell-1$:
\beq
b_\ell^{(NA)} = \sum_{k=0}^{\ell-1} b_{\ell,k}^{(NA)} \, (T_fN_f)^k \quad 
{\rm for} \ \ell \ge 2 \ . 
\label{bellk_nagt}
\eeq
Eq. (\ref{bellk_nagt}) is the analogue, for this non-Abelian gauge theory, of
the expansion (\ref{bellexpansion}) for the U(1) gauge theory.  Given that the
$b_\ell$ coefficients for the U(1) theory can be derived from those for the
non-Abelian theory by the formal replacements $C_A=0$, $C_f=1$, and $T_f=1$,
together with replacements of other group invariants that enter at loop level
$\ell \ge 4$ \cite{b4}, it follows that the $b_{\ell,0}$ terms vanish for the
U(1) theory.

Analogously to the U(1) theory, the large-$N_f$ dependence of the 
coefficients $b_\ell^{(NA)}$ in $\beta_\alpha^{(NA)}$
motivates the definition of rescaled coefficients that have finite limits as
$N_f \to \infty$, namely, for $\ell=1$, $b_1^{(NA)}/(T_fN_f) = 
b_{1,0}^{(NA)}/(T_fN_f) + b_{1,1}^{(NA)}$ and 
\beq
\check b_\ell^{(NA)} \equiv \frac{b_\ell^{(NA)}}{(T_fN_f)^{\ell-1}} 
\quad {\rm for} \ \ell \ge 2 \ . 
\label{bellck_nagt}
\eeq
We define 
\beq
\acute b_\ell^{(NA)} \equiv \frac{\check b_\ell^{(NA)}}{b_{1,1}^{(NA)}} 
= \frac{3}{4} \check b_\ell^{(NA)} 
\label{bellacute_nagt}
\eeq
and
\beq
\acute b_{\ell,k}^{(NA)} \equiv \frac{b_{\ell,k}^{(NA)}}{b_{1,1}^{(NA)}} 
= \frac{3}{4} b_{\ell,k}^{(NA)} \ . 
\label{bellkacute_nagt}
\eeq

Given that, for large $N_f$, $b_1^{(NA)} \sim (T_fN_f)$ and, for $\ell \ge 2$,
$b_\ell \sim (T_fN_f)^{\ell-1}$, one may construct a
rescaled beta function $\beta_\eta^{(NA)}$ that is finite in the LNF limit 
(\ref{lnf_nagt}) by defining 
\beq
\beta_\eta^{(NA)} = T_f N_f \beta_\alpha \ . 
\eeq
We have
\beq
\beta_\alpha^{(NA)} = 8 \pi a^2 \bigg [ b_{1,1}^{(NA)}T_fN_f + b_{1,0}^{(NA)} +
  \sum_{\ell=2}^\infty \check b_{\ell}^{(NA)} \, \eta^{\ell-1} \bigg ] \ . 
\label{betamixed_nagt}
\eeq
Thus, 
\beqs
\beta_\eta^{(NA)} & = & 8\pi b_{1,1}^{(NA)} \eta^2 \bigg [ 1 +
  \frac{\acute b_{1,0}^{(NA)}}{T_fN_f} + \frac{1}{T_fN_f} \sum_{\ell=2}^\infty 
\acute b_\ell^{(NA)} \, \eta^{\ell-1} \bigg ] \ . \cr\cr
& & 
\label{betaeta_nagt}
\eeqs
The $n$-loop rescaled beta function, $\beta_{\eta,n\ell}^{(NA)}$, is given by
Eq. (\ref{betaeta_nagt}) with the upper limit on the summation over loop order
$\ell$ set to $\ell=n$ rather than $\ell=\infty$.


\subsection{Possible UV Zero of Beta Function in a Non-Abelian 
Gauge Theory for Large $N_f$}

The condition that the $n$-loop beta function vanishes away from the
origin $\eta=0$ is the polynomial equation (of degree $n-1$ in $\eta$) 
\beq
1 + \frac{\acute b_{1,0}^{(NA)}}{T_fN_f} +  \frac{1}{T_fN_f} 
\sum_{\ell=2}^n \acute b_\ell^{(NA)} \, \eta^{\ell-1} = 0 \ . 
\label{betayzero_nagt}
\eeq
By an analysis similar to the one give above for the U(1) theory, in the LNF
limit (\ref{lnf_nagt}), of the $n-1$ roots of Eq.  (\ref{betayzero_nagt}), the
relevant one has the approximate form
\beq 
\eta_{_{UV,n\ell}} \sim
\bigg ( -\frac{T_fN_f}{\acute b_{n,n-1}^{(NA)}} \bigg )^{\frac{1}{n-1}} = 
\bigg ( -\frac{b_{1,1}^{(NA)}T_fN_f}{b_{n,n-1}^{(NA)}} 
\bigg )^{\frac{1}{n-1}} \ . 
\label{etasol}
\eeq
Since $b_1^{(NA)}$ and (for NFI STs) $b_{n,n-1}^{(NA)}$ are scheme-independent,
this solution for $\eta_{_{UV,n\ell}}$ in Eq. (\ref{etasol})is also NFI
scheme-independent. From Eq. (\ref{etasol}), it follows that a (NFI
scheme-independent) criterion for existence of a UV zero in the beta function
of this non-Abelian gauge theory for large $N_f$ is:
\beq
{\rm As} \ N_f \to \infty, \ \beta_{y,n\ell}^{(NA)} \ {\rm has \ a \ 
UV \ zero \ if \ and \ only \ if } \ b_{n,n-1}^{(NA)} < 0 \ .
\label{criterion_nagt}
\eeq
Let us assume that for some $n$, $b_{n,n-1}^{(NA)} < 0$, so
that $\beta_{\alpha,n\ell}^{(NA)}$ does have a UV zero for large $N_f$.  From
Eq. (\ref{etasol}), it follows that, for fixed loop order $n$ and
$N_f \to \infty$,
\beq
\lim_{N_f \to \infty} \eta_{_{UV,n\ell}} = \infty \ .
\label{etalnflimit}
\eeq
This is the analogue, for the non-Abelian gauge theory, of
Eq. (\ref{ylnflimit}) for the U(1) gauge theory.

We can express the rescaled beta function as 
\beq
\beta_\eta^{(NA)} = 8\pi b_{1,1}^{(NA)} \, \eta^2  
\bigg [ 1 + \sum_{s=1}^\infty \frac{F_s^{(NA)}(\eta)}{(T_f N_f)^s} \bigg ] 
\ . 
\label{betay_nagt}
\eeq
As discussed above, $F_1^{(NA)}(\eta)$ is NFI scheme-independent, while the
$F_s^{(NA)}(\eta)$ with $s \ge 2$ are scheme-dependent. The $F_s^{(NA)}(\eta)$
are expanded as
\beq
F_s^{(NA)}(\eta) = \sum_{p=s-1}^\infty f_{s,p}^{(NA)} \, \eta^p \ . 
\label{sfp_nagt}
\eeq

From the results in \cite{graceybeta}, the following closed-form
expression has been inferred for $F_1(\eta)^{(NA)}$ \cite{holdom2010}: 
\beq
F_1^{(NA)}(\eta) = -\frac{11C_A}{4} + \int_0^{4\eta/3} I_1(x)I_2(x) \, dx \ , 
\label{f1_nagt}
\eeq
where $I_1(x)$ was given above and 
\beq
I_2(x) = C_f + \frac{(20-43x+32x^2-14x^3+4x^4)C_A}{4(1-x^2)(2x-1)(2x-3)} \ . 
\label{i2_nagt}
\eeq
The function $I_2(x)$ has a simple pole at $|x|=1$.  Consequently, the
integral, $F_1^{(NA)}(\eta)$ is  analytic about $\eta=0$ in a disk in the
complex $\eta$ plane of radius $|\eta|=3/4$.  As $\eta$ approaches the value
\beq
\eta_{div} = \frac{3}{4} 
\label{etadiv}
\eeq
from below, $F_1^{(NA)}(\eta)$ diverges through negative values. Thus, as in
the U(1) case, this leads to a zero in the expression for $\beta_\eta^{(NA)}$
to leading order in $1/N_f$, i.e., there exists a value of $\eta$ slightly less
than 3/4 for which $[1+F_1^{(NA)}(\eta)/N_f]=0$ for large $N_f$.

Via Taylor series expansion of the integrand in Eq. (\ref{f1_nagt}) and
integration term-by-term, one can calculate the $f_{s,p}^{(NA)}$ and hence the
$b_{\ell,\ell-1}^{(NA)}$.  The $b_{\ell,\ell-1}^{(NA)}$ obtained in this way
agree with the known results for loop order $\ell=1$ to $\ell=4$. For $\ell=1$,
\beq
b_{1,1}^{(NA)} = \grave f_{1,0}^{(NA)} = b_{1,1} = \frac{4}{3} \ . 
\label{b11_nagt}
\eeq
For $\ell \ge 2$, the $b_{\ell,\ell-1}^{(NA)}$ have the general form 
\beq
b_{\ell,\ell-1}^{(NA)} = b_{\ell,\ell-1} \, C_f + b_{\ell,\ell-1,C_A} \, C_A 
\ , 
\label{bell_nagt_form}
\eeq
where $b_{\ell,\ell-1}$ is the corresponding coefficient for the U(1) gauge
theory. Specifically, 
\beqs
b_{2,1}^{(NA)} = \grave f_{1,1}^{(NA)} & = & b_{2,1} \, C_f + \frac{20}{3} C_A 
= 4C_f + \frac{20}{3}C_A \ , 
\cr\cr
& & 
\label{b21_nagt}
\eeqs
\beqs
b_{3,2}^{(NA)} = \grave f_{1,2}^{(NA)} & = & 
b_{3,2} \, C_f - \frac{158}{27}C_A \cr\cr
& = & - \bigg [ \frac{44}{9}C_f + \frac{158}{27}C_A \bigg ] \cr\cr
  & = & -(4.88889C_f + 5.85185C_A) \cr\cr
& & 
\label{b32_nagt}
\eeqs
and
\beqs
b_{4,3}^{(NA)} = \grave f_{1,3}^{(NA)} & = & 
b_{4,3} \, C_f - \frac{424}{243}C_A \cr\cr
& - & \bigg [ \frac{1232}{243}C_f + \frac{424}{243}C_A \bigg ] \cr\cr 
& = & -(5.06996C_f + 1.74486C_A) \ . 
\cr\cr
& & 
\label{b43_nagt}
\eeqs
The coefficients up to $\ell=7$ are \cite{graceybeta} 
\beqs
b_{5,4}^{(NA)} = \grave f_{1,4}^{(NA)} & = &  
b_{5,4} \, C_f + \Big ( -\frac{916}{243}+\frac{640}{81} \Big )C_A  \cr\cr
& = & 9.22127C_f + 5.72819C_A
\label{b54_nagt}
\eeqs
\begin{widetext}
\beqs
b_{6,5}^{(NA)} = \grave f_{1,5}^{(NA)} & = &  
b_{6,5} \, C_f + \Big ( -\frac{4832}{1215} - \frac{40448\zeta(3)}{3645} + 
\frac{512\pi^4}{6075} \Big )C_A \cr\cr
& = & 1.47275C_f - 3.63329C_A 
\label{b65_nagt}
\eeqs
and
\beqs
b_{7,6}^{(NA)} = \grave f_{1,6}^{(NA)} & = &
b_{7,6} \, C_f + \Big ( -\frac{9440}{2187}-\frac{27136\zeta(3)}{6561}
-\frac{20224\pi^4}{98415} + \frac{20480\zeta(5)}{729} \Big ) C_A  \cr\cr
& = & -(7.80879C_f + 0.1746567C_A) \ . 
\label{b76_nagt}
\eeqs
\end{widetext}

For the purpose of our analysis of zeros of the beta function of the general
non-Abelian theory in the large-$N_f$ limit, we have 
calculated the $b_{n,n-1}^{(NA)}$ to higher-loop order. For $\ell = 8$ we 
find
\begin{widetext}
\beqs
b_{8,7}^{(NA)} = \grave f_{1,7}^{(NA)} & = & b_{8,7} \, C_f + 
\Big ( -\frac{215680}{45927} - \frac{468992\zeta(3)}{45927} 
-\frac{54272\pi^4}{688905} - \frac{647168\zeta(5)}{15309} 
+ \frac{163840\pi^6}{2893401} + \frac{163840\zeta(3)^2}{15309} \Big ) C_A 
\cr\cr
& = & 1.42326C_f + 2.49243C_A \ . 
\label{b87_nagt}
\eeqs
\end{widetext}
In Appendix \ref{f1taylor} and Table \ref{bellnfmax_nagt} we list our
additional results for the $b_{\ell,\ell-1}^{(NA)}$ with two higher values of 
$\ell \equiv n$. 

In accordance with (\ref{criterion_nagt}), the $n$-loop beta function,
$\beta_{\alpha,n\ell}^{(NA)}$, has a UV zero for large $N_f$ if and only if
$b_{n,n-1}^{(NA)} < 0$.  We list the $b_{n,n-1}^{(NA)}$ in Table 
\ref{bellnfmax_nagt} together with an entry indicating whether or not the
$n$-loop beta function, $\beta_{\alpha,n\ell}^{(NA)}$ has a UV zero. 
The signs of the $b_{n,n-1}^{(NA)}$ are definite for
$b_{2,1}^{(NA)}$ (positive), $b_{3,2}^{(NA)}$ (negative), $b_{4,3}^{(NA)}$
(negative), and $b_{5,4}^{(NA)}$ (positive). 

In the case of $b_{6,5}^{(NA)}$ and some $b_{n,n-1}^{(NA)}$ for higher $n$, the
$C_f$ and $C_A$ terms have opposite signs, so that further analysis is
necessary. Let us define the ratio
\beq
r_C \equiv \frac{C_A}{C_f} \ . 
\label{rc}
\eeq
The ranges of values of $r_C$ for various fermion representations $R$ will be
relevant for our analysis and are given in Appendix \ref{rcvalues}. 
Analytically continuing $r_C$ from ${\mathbb Q}$ to ${\mathbb R}$, we find that
$b_{6,5}^{(NA)}$ is negative (positive) for $r_C$ greater (less) than
\beq
(r_C)_{6,5} = 
\frac{2(1255-2640\zeta(3)+24\pi^4)}{5(453+1264\zeta(3)-16\pi^4)} = 0.405348
\label{rc_b65_nagt_zero}
\eeq
One can reexpress this as a correlated condition on the non-Abelian gauge
group, $G$, and fermion representation, $R$. For example, consider $G={\rm
SU}(N_c)$, and $R$ equal to the fundamental representation. Then (see Appendix
\ref{rcvalues}) $r_C$ decreases from 8/3 at $N_c=2$ toward 2 as $N_c \to
\infty$, so that the condition that $(r_C)_{6,5}$ be greater than 0.405348 is
always satisfied, and hence $\beta_{\alpha,6\ell}^{(NA)}$ has a UV zero for
large $N_f$.  All of the other representations considered in Appendix
\ref{rcvalues}, including the adjoint, rank-2 symmetric ($S_2$), and rank-2
antisymmetric ($A_2$) tensor representations yield $r_C$ values larger than the
value in Eq. (\ref{rc_b65_nagt_zero}), and hence $\beta_{\alpha,6\ell}^{(NA)}$
also has a UV zero for large $N_f$ for all of these representations. 

Proceeding to higher-loop values, the coefficient $b_{7,6}^{(NA)}$ is
negative-definite, so that $\beta_{\alpha,7\ell}^{(NA)}$ has a UV zero for
large $N_f$.  However, the sign of $b_{8,7}^{(NA)}$ is positive, so
$\beta_{\alpha,8\ell}$ does not have a UV zero.  The signs of several of the
$b_{n,n-1}^{(NA)}$ with $n \ge 9$ depend on $r_C$. As an example, the sign of
$b_{9,8}^{(NA)}$ is negative if and only if $r_C > 2.2970$ (to the indicated
floating-point accuracy). Whether or not this condition is satisfied depends on
the gauge group and the fermion representation.  For definiteness, we again
focus on the group $G={\rm SU}(N_c)$.  If the fermions are in the fundamental
representation, then the condition $r_C > 2.2970$ corresponds to the condition
that $N_c$ (analytically continued from positive integer to positive real
values) is less than 2.7810. Thus, for this fermion representation and
(physical, integer) values of $N_c$, $b_{9,8}^{(NA)}$ is negative if and only
if $N_c=2$, and for this single value of $N_c$, $\beta_{\alpha,9\ell}^{(NA)}$
has a UV zero for large $N_f$. If the theory contains (Dirac) fermions in the
adjoint representation, then $r_C=1$, so $b_{9,8}^{(NA)}$ is positive.
Similarly, if the fermions are in the $S_2$ representation, then $r_C$ is
bounded between 8/9 and 1, so again, $b_{9,8}^{(NA)}$ is positive.  If the
fermions are in the $A_2$ representation, then the condition $r_C > 2.2970$
corresponds to the inequality $N_c < 2.965$. Since the $A_2$ representation is
defined only for $N_c \ge 3$, there is no physical $N_c$ that satisfies this
inequality. Thus, for the adjoint, $S_2$, and $A_2$ representations,
$\beta_{\alpha,9\ell}^{(NA)}$ has no UV zero for large $N_f$.

At loop order $n=11$, the condition for $b_{11,10}^{(NA)}$ to be negative is
that $r_C < 0.450206$, but this is never satisfied for any of the
representations considered here.  For these, $b_{11,10}^{(NA)}$ is positive, so
$\beta_{\alpha,11\ell}$ does not have any UV zero for large $N_f$.  At loop
order $n=12$, the condition for $b_{12,11}^{(NA)}$ to be negative is that $r_C
> 1.3528$.  This inequality is always satisfied for the fundamental
representation, but is never satisfied for the adjoint, or $S_2$
representations. For the $A_2$ representation, this inequality corresponds to
the inequality $N_c < 5.28542$.  Hence, for the physical, integer values
$N_c=3, \ 4, \ 5$, this inequality is satisfied, so that $b_{12,11}^{(NA)} < 0$
and $\beta_{\alpha,11\ell}^{(NA)}$ has a UV zero at large $N_f$.

Up to loop order $n=18$, there are two other cases where the sign of
$b_{n,n-1}^{(NA)}$ depends on $r_C$, but in both of these cases, namely $n=15$
and $n=18$, the respective inequalities for $b_{n,n-1}^{(NA)}$ to be negative
are satisfied for all of the representations that we consider, so that for
$n=13$ up to this highest loop order, $n=18$, the respective
$\beta_{\alpha,n\ell}^{(NA)}$ has a UV zero for large $N_f$. 

It is thus interesting that for the seven cases $n=12$ through $n=18$,
$b_{n,n-1}^{(NA)}$ is negative for $R$ equal to the fundamental representation.
This is to be contrasted with the U(1) theory, in which some of the $b_{n,n-1}$
in this interval of $n$ (specifically, $n=12, \ 15, \ 18$) are positive.
However, even in the cases where $b_{n,n-1}^{(NA)} < 0$, so that there is a
solution for $\eta_{_{UV,n\ell}}$, it has the property (\ref{etalnflimit}).  In
this respect, the situation with finite-loop calculations of a UV zero in the
beta function for a non-Abelian gauge theory for large $N_f$ is similar to the
situation with the Abelian theory.

As is evident from Eq. (\ref{f1_nagt}), $F_1^{(NA)}(\eta)$ is equal to
$-(11/4)C_A$ at $\eta=0$.  As a function of $\eta$, it increases as $\eta$
increases from zero through negative values, reaching a broad maximum at $\eta
\simeq 0.6$ and then decreases again and diverges logarithmically through
negative values as $\eta$ approaches 3/4 from below. We will focus on the
interval
\beq
I_\eta: \quad 0 \le \eta \le \frac{3}{4} \ .
\label{etainterval}
\eeq
in our analysis. The condition that $\beta_\eta^{(NA)}$, calculated to
leading order in $1/(T_fN_f)$, vanishes with a UV zero is 
\beq
1+\frac{F_1^{(NA)}(\eta)}{T_fN_f} = 0 \ . 
\label{betay_nagt_zero_leading}
\eeq
Since $F_1^{(NA)}(\eta)$ diverges negatively as $\eta$ approaches 3/4 from
below, there exists a value of $\eta$ slightly less than 3/4 for which the
condition (\ref{betay_nagt_zero_leading}) is satisfied.  This suggests that the
rescaled beta function $\beta_\eta^{(NA)}$ might have a UV zero. As in the
Abelian case, the contributions of higher-order terms $\sum_{s=2}^\infty
F_s^{(NA)}(\eta)/(T_fN_f)^s$ might be such that the full beta function does not
have a UV zero that can be reached via renormalization-group evolution from
small couplings in the infrared. 

As we did for the U(1) theory, we may compare these results with findings from
analyses of possible UV zeros in the $n$-loop beta function
$\beta_{\alpha,n\ell}^{(NA)}$ for fixed large $N_f$.  The criterion
(\ref{criterion_nagt}) determines whether such a UV zero exists for large
$N_f$.  We have found a scatter of signs in the $b_{n,n-1}^{(NA)}$, as listed
in Table \ref{bellnfmax_nagt}, although the $b_{n,n-1}^{(NA)}$ for $n=12$ to
$n=18$ are negative for fermions in the fundamental representation.  Even for
the values of $n$ for which there is a UV zero, the solution obeys the
asymptotic limiting relation (\ref{etalnflimit}), making it difficult to match
with a finite value of $\eta$ for a possible UV zero in $\beta_\eta^{(NA)}$.


\section{Conclusions}
\label{conclusions}

In this paper we have extended our earlier study \cite{sch2} of a UV zero in
the $n$-loop beta function, $\beta_{\alpha,n\ell}$ in a U(1) gauge theory to
large values of $N_f$ and have investigated how the results relate to exact
results for the leading correction term $F_1(y)/N_f$, in the (rescaled) beta
function $\beta_y$ obtained by summation of a certain dominant set of Feynman
diagrams up to infinitely high loop order in the LNF limit (\ref{lnf}).
Effects of scheme transformations on the $b_{\ell,k}$ and $f_{s,p}$ have been
calculated.  A general criterion was given for determining whether or not the
$n$-loop $\beta$ function has a UV zero for large $N_f$, namely that
$b_{n,n-1}$, which is NFI scheme-independent, must be negative. As part of our
study, we have presented new analytic and numerical results for the
coefficients $b_{n,n-1}$ that enter as leading-$N_f$ terms in the $n$-loop
coefficients $b_n$ in the beta function. The coefficients $b_{n,n-1}$ show a
scatter of both positive and negative values and hence do not give evidence for
a stable UV zero in the U(1) beta function at large $N_f$ up to the
highest-loop order, namely $n=24$, to which we probed.  We derived, and
verified the accuracy of, an approximate analytic expression (\ref{ysol}) for
the UV zero of the $n$-loop beta function and showed that, even if $b_{n,n-1} <
0$ so that the $n$-loop beta function has a UV zero, the value of
$y_{_{UV,n\ell}}$ diverges as $N_f \to \infty$. By calculating corresponding
coefficients $b_{n,n-1}^{(d)}$ arising from the negative logarithmically
divergent term in $F_1(y)$ and determining how these differ from the actual
$b_{n,n-1}$, we showed that the latter, at least to loop order $n=24$, are not
sensitive to the negatively divergent term in $F_1(y)$. We analyzed various
illustrative test functions for $\beta_y$ incorporating a UV zero and
calculated resultant series expansions in $1/N_f$ and in $y$ for these
functions, finding that the resultant $b_{n,n-1}$ are uniformly negative, as is
also true of the $b_{n,n-1}^{(d)}$.

We have also considered the analogous question of a UV zero in the $n$-loop
beta function in a non-Abelian gauge theory at large $N_f$. We have shown that
for some loop orders $n$, ${\rm sgn}(b_{n,n-1})$, and hence the existence of a
UV zero in the $n$-loop beta function at large $N_f$, can depend on the fermion
representation, $R$, and have discussed the consequences of the ranges of
values of the relevant ratio $r_C \equiv C_A/C_f$ for various representations.
In general, our conclusions for the non-Abelian gauge theory are broadly
similar to those that we reach for the U(1) theory.  It is hoped that these
results will be a useful addition to the understanding of the properties of the
beta functions of Abelian and non-Abelian gauge theories.


\begin{acknowledgments}

I would like to thank Bob Holdom for valuable discussions and comments.  This
research was partly supported by the NSF grants NSF-PHY-09-69739 and
NSF-PHY-13-16617.

\end{acknowledgments}

\bigskip
\bigskip


\begin{appendix}

\section{Effects of Scheme Transformations on $b_{\ell,k}$ and $f_{s,p}$ } 
\label{nfstappendix}

\subsection{Transformations of $b_{\ell,k}$} 

In Sect. \ref{nfst} we gave a general formula for the effect of NFI scheme
transformations on the coefficients $b_{\ell,k}$ multiplying $N_f^k$ in the
$\ell$-loop term, $b_\ell$ in the beta function of a U(1) gauge theory Here we
list some of these relations explicitly for $3 \le \ell \le 6$.  In addition to
the invariance relation (\ref{bnnminus1invariance}), we have
\beqs
b_{3,1}' & = & b_{3,1} + h_{3,2}b_{2,1} + h_{3,1}b_{1,1} \cr\cr
         & = & b_{3,1} +    k_1 b_{2,1} + (k_1^2-k_2)b_{1,1} \ , 
\label{b31prime}
\eeqs
\beq
b_{4,2}' = b_{4,2} + h_{4,3} b_{3,2} = b_{4,2} + 2k_1 b_{3,2} \ , 
\label{b42prime}
\eeq
\beqs
b_{4,1}' & = & b_{4,1} + h_{4,3}b_{3,1} + h_{4,2}b_{2,1} + h_{4,1}b_{1,1}
\cr\cr
         & = & b_{4,1} + 2k_1 b_{3,1} + k_1^2 b_{2,1} + 
2(-k_1^3+2k_1k_2-k_3)b_{1,1} \ , \cr\cr
& & 
\label{b41prime}
\eeqs
\beq
b_{5,3}' = b_{5,3} + h_{5,4}b_{4,3} = b_{5,3}   + 3k_1 b_{4,3} \ , 
\label{b53prime}
\eeq
\beqs
b_{5,2}' & = & b_{5,2} + h_{5,4}b_{4,2} + h_{5,3} b_{3,2} \cr\cr
         & = & b_{5,2} + 3k_1 b_{4,2} + (2k_1^2+k_2)b_{3,2} \ , 
\label{b52prime}
\eeqs
\beqs
b_{5,1}' & = & b_{5,1} + h_{5,4}b_{4,1} + h_{5,3}b_{3,1} + h_{5,2}b_{2,1} 
+ h_{5,1}b_{1,1} \cr\cr
         & = &  b_{5,1} + 3k_1 b_{4,1} + (2k_1^2+k_2)b_{3,1} \cr\cr
         & + & (-k_1^3+3k_1k_2-k_3)b_{2,1} \cr\cr
& + & (4k_1^4-11k_1^2k_2+6k_1k_3+4k_2^2-3k_4)b_{1,1} \ , \cr\cr
& & 
\label{b51prime}
\eeqs
\beq
b_{6,4}' = b_{6,4} + h_{6,4}b_{5,4} = b_{6,4} + 4k_1 b_{5,4} \ , 
\label{b64prime}
\eeq
\beqs
b_{6,3}' & = & b_{6,3} + h_{6,5}b_{5,3} + h_{6,4}b_{4,3} \cr\cr
         & = & b_{6,3} + 4k_1 b_{5,3} + 2(2k_1^2+k_2)b_{4,3}  \ , 
\label{b63prime}
\eeqs
\beqs
b_{6,2}' & = & b_{6,2}+h_{6,5}b_{5,2}+h_{6,4}b_{4,2}+h_{6,3}b_{3,2} \cr\cr
         & = & b_{6,2} + 4k_1 b_{5,2} + 2(2k_1^2+k_2)b_{4,2} + 4k_1k_2 b_{3,2} 
\ , \cr\cr
& & 
\label{b62prime}
\eeqs
and
\begin{widetext}
\beqs
b_{6,1}' & = & b_{6,1}+h_{6,5}b_{5,1}+h_{6,4}b_{4,1}+h_{6,3}b_{3,1}+
h_{6,2}b_{2,1}+h_{6,1}b_{1,1} \cr\cr
         & = & b_{6,1} + 4k_1 b_{5,1} + 2(2k_1^2+k_2)b_{4,1} + 4k_1k_2 b_{3,1}
+ (2k_1^4-6k_1^2k_2+4k_1k_3+3k_2^2-2k_4)b_{2,1} \cr\cr
& + & 4(-2k_1^5+7k_1^3k_2-4k_1^2k_3-5k_1k_2^2+2k_1k_4+3k_2k_3-k_5)b_{1,1} \ , 
\label{b61prime}
\eeqs
\end{widetext}
and similarly for $\ell \ge 7$.  These formulas determine the
corresponding transformations of the $f_{s,p}$ in Eqs. (\ref{fsypexpansion})
and (\ref{sfp_nagt}).  Similar results hold for a non-Abelian gauge theory. 

\subsection{Transformations of $f_{s,p}$ and $F_s(y)$ }

Here we present calculations of the $f_{s,p}'$ in terms of $f_{s,p}$ for NFI
scheme transformations, and the resultant transformation of $F_s(y)$, for $s=2,
\ 3$.  Since for $\ell \ge 3$, the $b_{\ell,k}'$ are not, in general, equal to
the $b_{\ell,k}$ for the lower values of $k$, $1 \le k \le \ell-2$, it follows
that the $f_{s,p}$ and $F_s(y)$ with $s \ge 2$ are scheme-dependent even for
NFI STs.  Using Eqs. (\ref{fsp_to_bellk}) or equivalently, (\ref{bellk_to_fsp})
together with our general formulas for the scheme transformation of
$b_{\ell,k}$ with $1 \le k \le \ell-2$, Eqs. (\ref{bellkprimebellk}) and
(\ref{bell1primebell1}), we can calculate how the $f_{s,p}$ and hence $F_s(y)$
with $s \ge 2$ change under an NFI scheme transformation.  For the coefficients
$f_{2,p}$ that occur in the $s=2$ special case of Eq. (\ref{fsypexpansion}), we
find
\beqs
f_{2,2}' & = & f_{2,2} + h_{3,2}f_{1,1} + h_{3,1} \cr\cr
         & = & f_{2,2} +    k_1 f_{1,1} + (k_1^2-k_2) \ , 
\label{f22prime}
\eeqs
and
\beqs
f_{2,p}' & = & f_{2,p} + h_{p+1,p}f_{1,p-1} \cr\cr
         & = & f_{2,p} + (p-1)k_1 f_{1,p-1} \quad {\rm for} \ p \ge 3 \ . 
\label{f2pprimepge3}
\eeqs
Substituting these results into the $s=2$ special case of
Eq. (\ref{fsypexpansion}), 
we thus derive the effect of an NFI scheme transformation on $F_2(y)$:
\beqs
F_2(y)_{ST} & = & F_2(y) + (h_{32}f_{1,1}+h_{3,1})y^2 + 
\sum_{p=3}^\infty h_{p+1,p} \, f_{1,p-1} \, y^p \cr\cr
            & = & F_2(y) + (k_1 f_{1,1}+k_1^2-k_2)y^2 \cr\cr
            & + & k_1 \sum_{p=3}^\infty (p-1)f_{1,p-1} \, y^p \ . 
\label{f2st}
\eeqs

For $s=3$ we find 
\beqs
f_{3,3}' & = & f_{3,3} + h_{4,3}f_{2,2} + h_{4,2}f_{1,1} + h_{4,1} \cr\cr
         & = & f_{3,3} + 2k_1 f_{2,2} + k_1^2 f_{1,1} + (-2k_1^3+4k_1k_2-2k_3) 
\cr\cr
& & 
\label{f33prime}
\eeqs
and
\beq
f_{3,p}' = f_{3,p} + h_{p+1,p}f_{2,p-1} + h_{p+1,p-1}f_{1,p-2} 
\quad {\rm for} \ p \ge 4 \ .
\label{f3pprimepge4}
\eeq
Hence, 
\beqs
F_3(y)_{ST} & = & F_3(y) + (h_{4,3}f_{2,2}+h_{4,2}f_{1,1}+h_{4,1})y^3 \cr\cr
& + & \sum_{p=4}^\infty (h_{p+1,p}f_{2,p-1} + h_{p+1,p-1}f_{1,p-2})y^p \ . 
\cr\cr
& & 
\label{f3st}
\eeqs
Similar results follow for $s \ge 4$. 


\section{Some Relevant Properties of the Euler Gamma Function}
\label{gammafunction}

We recall the definition of the Euler gamma function $\Gamma(x)$:
\beq
\Gamma(x) = \int_0^\infty e^{-t} \, t^{x-1} \, dt \ . 
\label{Gamma}
\eeq
The function $\Gamma(x)$ has simple poles at $x=-n+\epsilon$, with $n=0, \ 1, \
2, ...$. The Taylor-Laurent series for $\Gamma(x)$ as $x$ approaches one of
these poles is
\beq
\Gamma(-n+\epsilon) \sim \frac{(-1)^n}{n!}\Big [ \frac{1}{\epsilon} + 
\Big ( -\gamma_E + \sum_{m=1}^n \frac{1}{m} \Big ) + O(\epsilon) \Big ] \ . 
\label{eulergammapole}
\eeq
where the term $\sum_{m=1}^n \frac{1}{m}$ is absent for $n=0$ and $\gamma_E$ is
the Euler-Mascheroni constant, 
\beq
\gamma_E = \lim_{k \to \infty} \Big [ \Big (\sum_{s=1}^k \frac{1}{s} \Big ) 
- \ln k \Big ] = 0.5772... 
\label{gammae}
\eeq

The expansion of $I_1(x)$ about $x=0$ and hence $F_1(y)$ about $y=0$ makes use
of the Taylor series expansions of the gamma functions in $I_1(x)$ about
nonsingular points.  These in turn rely upon the basic results (e.g.,
\cite{zeta}) 
\beq
\Gamma(x+1) = \sum_{k=0}^\infty c_k x^k \ ,
\label{gammataylor}
\eeq
where $c_0=1$ and
\beq
c_{n+1} = (n+1)^{-1}\sum_{m=0}^n (-1)^{m+1} s_{m+1}c_{n-m}
\label{cn}
\eeq
with
\beq
s_1 = \gamma_E, \quad s_m = \zeta(m) \ {\rm for} \ m \ge 2
\label{s1n}
\eeq
and 
\beq
\frac{1}{\Gamma(x+1)} = \sum_{k=0}^\infty d_k x^k \ , 
\label{gammainversetaylor}
\eeq
where $d_0=1$ and
\beq
d_{n+1} = (n+1)^{-1}\sum_{m=0}^n (-1)^m c_{m+1}d_{n-m} \ . 
\label{dn}
\eeq
These relations explain the presence of the Riemann $\zeta(s)$ functions in
various $f_{s,p}$ and $b_{n,k}$. As noted in the text, the $\zeta(m)$ with even
$m=2r$ are evaluated using the identity
\beq
\zeta(2r)=\frac{(-1)^{r+1}B_{2r}(2\pi)^{2r}}{2(2r)!} \ , 
\label{zetaeven}
\eeq
where the $B_n$ are the Bernoulli numbers, defined by 
\beq
\frac{t}{e^t-1} = \sum_{n=0}^\infty B_n \frac{t^n}{n!} \ . 
\label{bernoulli}
\eeq
Thus, $B_0=1$, $B_1=-1/2$, $B_2=1/6$, $B_4=-1/30$, and $B_{2r+1}=0$ for $r \ge
1$.  Note that ${\rm sgn}(B_{2r})=(-1)^{r-1}$.  The relations for $\zeta(2r)$
are then 
\beq
\zeta(2)=\frac{\pi^2}{6} \ , \quad \zeta(4)=\frac{\pi^4}{90} \ , \quad 
\zeta(6)=\frac{\pi^6}{945} \ , 
\label{somezetas}
\eeq
and so forth for higher values of $2r$. 


\section{Taylor Series Expansions of $F_1(y)$ and $F_1^{(NA)}(\eta)$}
\label{f1taylor}

In this appendix we list the coefficients $f_{s,p}$ and $f_{s,p}^{(NA)}$ in the
respective Taylor series expansions of $F_1(y)$ and $F_1^{(NA)}(\eta)$ for the
U(1) and non-Abelian gauge theories. For the U(1) theory we have 
\beq
f_{1,1} = 3 \ , 
\label{f11}
\eeq
\beq
f_{1,2} = -\frac{11}{3} = -3.66667 \ , 
\label{f12} 
\eeq
\beq
f_{1,3} = -\frac{308}{81} = -3.80247 \ , 
\label{f13}
\eeq
\beq
f_{1,4} = \frac{214}{81} + \frac{32\zeta(3)}{9} = 6.915955 \ , 
\label{f14}
\eeq
\beq
f_{1,5} = \frac{4016}{1215}-\frac{2816\zeta(3)}{405}+
\frac{128\pi^4}{2025} = 1.10456 \ , 
\label{f15}
\eeq
\medskip

\beqs
f_{1,6} & = & \frac{1072}{243}-\frac{19712\zeta(3)}{2187}
-\frac{1408\pi^4}{10935}+\frac{1024\zeta(5)}{81} \cr\cr
& = & -5.85659 \ , 
\label{f16}
\eeqs
and so forth for higher $f_{s,p}$ coefficients. 

For the non-Abelian gauge theory we have 
\beq
f_{1,0}^{(NA)} = -\frac{11C_A}{4} \ , 
\label{f10_nagt}
\eeq
\beq
f_{1,1}^{(NA)} = 3C_f+5C_A \ , 
\label{f11_nagt}
\eeq
\bigskip

\beq
f_{1,2}^{(NA)} = - \bigg [ \frac{11C_f}{3} + \frac{79C_A}{18} \bigg ] \ , 
\label{f12_nagt} 
\eeq
\medskip
\beq
f_{1,3}^{(NA)} = - \bigg [ \frac{308C_f}{81} + \frac{106C_A}{81} \bigg ] \ , 
\label{f13_nagt}
\eeq
\beqs
f_{1,4}^{(NA)} & = & 
\Big ( \frac{214}{81}+\frac{32\zeta(3)}{9} \Big ) C_f \cr\cr
& + & \Big ( -\frac{229}{81} + \frac{160\zeta(3)}{27} \Big ) C_A \ , 
\label{f14_nagt}
\eeqs
and so forth for higher orders. 


\section{Higher-Loop Coefficients $b_{n,n-1}$ and $b_{n,n-1}^{(NA)}$ } 
\label{bnnminus1}

\subsection{U(1) Gauge Theory}

In the text, for the U(1) gauge theory with $N_f$ fermions we have listed the
$b_{n,n-1}$ up to $n=8$.  Here we list the higher-loop coefficients for $n=9, \
10$: 
\begin{widetext}
\beqs
b_{9,8} = \grave f_{1,8} 
& = & \frac{214720}{19683}+\frac{257024\zeta(3)}{19683}
+\frac{54784\pi^4}{295245}-\frac{315392\zeta(5)}{6561}
- \frac{90112\pi^6}{1240029}-\frac{90112\zeta(3)^2}{6561}
+\frac{16384\zeta(7)}{243}
+\frac{8192\pi^4\zeta(3)}{32805} \cr\cr
& = & 2.35202
\eeqs
and
\beqs
b_{10,9} = \grave f_{1,9} 
& = & \frac{2630656}{177147}+\frac{1097728\zeta(3)}{59049}
+\frac{2056192\pi^4}{7971615}
+\frac{7012352\zeta(5)}{177147}
-\frac{1441792\pi^6}{14348907} - \frac{10092544\zeta(3)^2}{531441} \cr\cr
& - & \frac{2883584\zeta(7)}{19683} -\frac{1441792\pi^4\zeta(3)}{2657205}
+\frac{229376\pi^8}{13286025} 
+\frac{1048576\zeta(3)\zeta(5)}{19683} \cr\cr
& = & -1.71453 \ . 
\label{b10_9}
\eeqs

\subsection{Non-Abelian Gauge Theory}

Similarly, we calculate 
\beqs
b_{9,8}^{(NA)} = \grave f_{1,8}^{(NA)} & = & b_{9,8} \, C_f + 
\Big ( -\frac{32992}{6561} - \frac{77312\zeta(3)}{6561} 
-\frac{58624\pi^4}{295245} - \frac{108544\zeta(5)}{6561} 
-\frac{323584\pi^6}{3720087} - \frac{323584\zeta(3)^2}{19683} \cr\cr
& + & \frac{81920\zeta(7)}{729} + \frac{8192\pi^4\zeta(3)}{19683} \Big ) C_A 
\cr\cr
& = & 2.35202C_f - 1.02394C_A
\label{b98_nagt}
\eeqs
and
\beqs
b_{10,9}^{(NA)} = \grave f_{1,9}^{(NA)} & = & b_{10,9} \, C_f +  
\Big ( -\frac{921088}{177147} - \frac{2416640\zeta(3)}{177147} 
-\frac{618496\pi^4}{2657205} - \frac{7503872\zeta(5)}{177147}
-\frac{3473408\pi^6}{100442349} - \frac{3473408\zeta(3)^2}{531441}
\cr\cr
& - & \frac{10354688\zeta(7)}{59049} -\frac{5177344\zeta(3)\pi^4}{7971615}
+\frac{229376\pi^8}{7971615} + \frac{5242880\zeta(3)\zeta(5)}{59049}
 \Big ) C_A 
\cr\cr
& = & - (1.71453C_f +0.059843C_A) \ . 
\label{b10comma9_nagt}
\eeqs
\end{widetext}
%


\section{Ranges of Values of $r_C$} 
\label{rcvalues}

The ranges of values of the ratio $r_C$ in Eq. (\ref{rc}) for various fermion
representations $R$ will be important for our analysis of possible UV zeros 
of the beta function in a non-Abelian gauge theory. For definiteness,
we focus on the gauge group $G={\rm SU}(N_c)$; it is straightforward to deal
with other gauge groups.  For $R$ equal to the
fundamental representation of SU($N_c$), one has
\beq
r_C = \frac{2N_c^2}{N_c^2-1} \quad {\rm for} \ R = {\rm fund.} 
\label{rc_fund}
\eeq
This has the value $r_C=8/3$ at $N_c=2$ and decreases monotonically with
increasing $N_c$, approaching the limiting value of 2 as $N_c \to
\infty$. Setting $r_C$ equal to a given
value and solving for $N_c$ yields a quadratic equation for $N_c$, with a
unique physical solution, 
\beq
N_c = \bigg ( \frac{r_C}{r_C-2} \bigg )^{1/2} \ . 
\label{ncrcsol_fund}
\eeq

We next consider two-index representations. For
$R$ equal to the adjoint representation, $r_C$ has the unique value
$r_C=1$. If $R$ is the symmetric rank-2 tensor representation, denoted
$S_2$, then 
\beq
r_C = \frac{N_c^2}{(N_c+2)(N_c-1)} \quad {\rm for} \ R = S_2 \ . 
\label{rc_sym2}
\eeq
This has the value 1 at $N_c=2$ and decreases to a minimum of 8/9 at
$N_c=4$, after which it increases for larger $N_c$, approaching the limiting
value of 1 from below as $N_c \to \infty$.  Regarding the inverse relation,
there is no solution if $r_C \not\in [8/9,1)$, and there are
two solutions for $N_c$ (as a formal real variable) if $r_C \in [8/9,1)$.
These solutions are 
\beq
N_c = \frac{r_C + \sqrt{r_C(9r_C-8)}}{2(1-r_C)} 
\label{ncrcsol1_sym2}
\eeq
and
\beq
N_c = \frac{r_C - \sqrt{r_C(9r_C-8)}}{2(1-r_C)} 
\label{ncrcsol2_sym2}
\eeq

If $R$ is the antisymmetric rank-2
tensor representation (defined for $N_c \ge 3$), denoted $A_2$, 
\beq
r_C = \frac{N_c^2}{(N_c-2)(N_c+1)} \quad  {\rm for} \ R = A_2 \ . 
\label{rc_asym2}
\eeq
This has the value $r_C=8/3$ for $N_c=3$ and decreases monotonically with
$N_c$, approaching the limiting value 1 from above as $N_c \to \infty$.

One can also consider more complicated representations. For example, if $R$ is
the symmetric rank-3 tensor representation, $S_3$, then
\beq
r_C = \frac{2N_c^2}{3(N_c+3)(N_c-1)} \quad {\rm for} \ R = S_3 \ . 
\label{rc_sym3}
\eeq
This has the value $r_C=8/15=0.5333$ at $N_c=2$ and increases with 
$N_c$, approaching the value 2/3 from below as $N_c \to \infty$. 
For $R$ equal to the antisymmetric rank-3 tensor representation $A_3$ (defined
for $N_c \ge 4$), 
\beq
r_C = \frac{2N_c^2}{3(N_c-3)(N_c+1)} \quad {\rm for} \ R = A_3 \ . 
\label{rc_asym3}
\eeq
This has the value $32/15 = 2.1333$ for $N_c=4$ and decreases monotonically
with $N_c$, approaching the limit 2/3 from above as $N_c \to \infty$.

\end{appendix}



\newpage

\begin{table}
\caption{\footnotesize{Values of the UV zero in the $\beta_\alpha$ function of
the U(1) gauge theory with $N_f$ fermions, at $n$-loop ($n\ell$) order, for
$n=3, \ 4, \ 5$, in the $\overline{\rm MS}$ scheme, denoted
$\alpha_{_{UV,n\ell}}$. The symbol $-$ indicates that there is no zero in
$\beta$ for the given order and value of $N_f$. See text for further details.}}
\begin{center}
\begin{tabular}{|c|c|c|c|} \hline\hline
$N_f$ & $\alpha_{_{UV,3\ell}}$ & $\alpha_{_{UV,4\ell}}$ &
$\alpha_{_{UV,5\ell}}$ 
\\ \hline 
1 & 10.2720 & 3.0400  & $-$      \\ 
2 & 6.8700  & 2.4239  & $-$      \\ 
3 & 5.3689  & 2.0776  & $-$      \\ 
4 & 4.5017  & 1.8463  & $-$      \\ 
5 & 3.9279  & 1.67685 & 2.5570   \\ 
6 & 3.5156 & 1.5455   & 1.8469   \\ 
7 & 3.2027 & 1.4397   & 1.6243   \\ 
8 & 2.9555 & 1.3519   & 1.4851   \\ 
9 & 2.7545 & 1.2776   & 1.3863   \\ 
10 & 2.5871 & 1.2135  & 1.3120   \\ 
20 & 1.7262 & 0.8483  & $-$      \\
100 & 0.7081 & 0.33265 & $-$     \\ 
500 & 0.3038 & 0.1203 & $-$      \\ 
$10^3$ & 0.2127 & 0.07678 & $-$  \\ 
$10^4$ & 0.016614 & 0.016965 & $-$ \\ 
\hline\hline
\end{tabular}
\end{center}
\label{alfuvu1}
\end{table}


\begin{table}
\caption{\footnotesize{Values of $\kappa_{n\ell}$ (defined in
Eq. (\ref{kappa_nloop})) as a function of $N_f$ in a U(1) gauge theory for
$n=3$ and $n=4$ loops.  See text for further details.}}
\begin{center}
\begin{tabular}{|c|c|c|} \hline\hline
$N_f$ & $\kappa_{3\ell}$ & $\kappa_{4\ell}$ 
\\ \hline
 1     &  1.8580  &  1.1249    \\
 2     &  1.6248  &  1.0772    \\
 3     &  1.5105  &  1.0490    \\
 4     &  1.4404  &  1.0297    \\
 5     &  1.3920  &  1.0155    \\
 6     &  1.3562  &  1.0045    \\
 7     &  1.3284  &  0.9957    \\
 8     &  1.3060  &  0.9885    \\
 9     &  1.2875  &  0.9825    \\
10     &  1.2719  &  0.9774    \\
20     &  1.1883  &  0.9515    \\ 
100    &  1.0812  &  0.9365    \\
500    &  1.0356  &  0.9512    \\
$10^3$ &  1.0251  &  0.9588    \\
$10^4$ &  1.0079  &  0.9786    \\
\hline\hline
\end{tabular}
\end{center}
\label{kappavalues} 
\end{table}


\begin{table}
\caption{\footnotesize{List of terms in $b_n$ of highest degree (equal to
degree $n-1$) in $N_f$ for $2 \le n \le 24$, denoted $b_{n,n-1}$ in a U(1)
gauge theory. The $n$-loop beta function in this theory,
$\beta_{\alpha,n\ell}$, has a UV zero at large $N_f$ if and only if $b_{n,n-1}$
is negative. Notation $a$e-m means $a \times 10^{-m}$. See text for
further details.}}
\begin{center}
\begin{tabular}{|c|c|} \hline\hline
$n$ & $b_{n,n-1}$    \\ \hline
 2  &    4           \\
 3  &  $-4.8889$     \\
 4  &  $-5.0700$     \\
 5  &    9.2213      \\
 6  &    1.47275     \\
 7  &  $-7.8088$     \\
 8  &    2.4924      \\
 9  &    2.3520      \\
10  &  $-1.7145$     \\
11  &  $-0.022326$   \\
12  &    0.37396     \\
13  &  $-0.108455$   \\
14  &  $-0.025088$   \\
15  &    0.020247    \\
16  &  $-2.2434$e-3  \\
17  &  $-1.3959$e-3  \\
18  &    0.49000e-3  \\
19  &    1.40184e-6  \\
20  &  $-0.32360$e-4 \\
21  &    0.59182e-5  \\
22  &    0.65388e-6  \\
23  &  $-0.39871$e-6 \\
24  &    0.33916e-7  \\
\hline\hline
\end{tabular}
\end{center}
\label{bellnfmax} 
\end{table}

\begin{table}
\caption{\footnotesize{List of coefficients $b_{n,n-1}^{(d)}$, as defined in 
Eq. (\ref{bnnminus1fdiv}) for $2 \le n \le 24$ in a U(1) gauge theory. 
Notation $a$e-m means $a \times 10^{-m}$. See text for further details.}}
\begin{center}
\begin{tabular}{|c|c|} \hline\hline
$n$ & $b_{n,n-1}^{(d)}$ \\ \hline
 2  &  $-0.033624$    \\
 3  &  $-0.89663$e-2  \\
 4  &  $-0.31880$e-2  \\
 5  &  $-1.2752$e-3   \\
 6  &  $-0.54409$e-3  \\
 7  &  $-0.24182$e-3  \\
 8  &  $-1.10545$e-4  \\
 9  &  $-0.51588$e-4  \\
10  &  $-0.24456$e-4  \\
11  &  $-1.1739$e-5   \\
12  &  $-0.56917$e-5  \\
13  &  $-0.27826$e-5  \\
14  &  $-1.3699$e-6   \\
15  &  $-0.67842$e-6  \\
16  &  $-0.33770$e-6  \\
17  &  $-1.6885$e-7   \\
18  &  $-0.84757$e-7  \\
19  &  $-0.42692$e-7  \\
20  &  $-0.21571$e-7  \\
21  &  $-1.0929$e-8   \\
22  &  $-0.55514$e-8  \\
23  &  $-0.28261$e-8  \\
24  &  $-1.4417$e-9   \\
\hline\hline
\end{tabular}
\end{center}
\label{bnminus1div}
\end{table}


\begin{table}
\caption{\footnotesize{List of terms in $b_n^{(NA)}$ of highest degree 
(equal to degree $n-1$) in $N_f$ for $2 \le n \le 18$, denoted 
$b_{n,n-1}^{(NA)}$, and resultant determination of the existence or 
non-existence of a UV zero of the $n$-loop
beta function, $\beta_{\alpha,n\ell}$, for large $N_f$ (entries 
Y and N denote yes and no). For some $n$, ${\rm sgn}(b_{n,n-1}^{(NA)})$ depends
on $r_C \equiv C_A/C_f$, as indicated. 
Notation $a$e-m means 
$a \times 10^{-m}$. See text for further discussion.}}
\begin{center}
\begin{tabular}{|c|c|c|} \hline\hline
$n$ & $b_{n,n-1}^{(NA)}$ &  $\exists \ \alpha_{_{UV,n\ell}}$ at large $N_f$? 
\\ \hline
 2  & $4C_f+6.6667C_A$              &  N  \\
 3  & $-(4.8889C_f+5.85185C_A)$     &  Y  \\
 4  & $-(5.0700C_f+1.7449C_A)$      &  Y  \\
 5  & $  9.2213C_f+5.7282C_A$       &  N  \\
 6  & $ 1.47275C_f-3.6333C_A$       &  Y  \ iff \ $r_C > 0.40535$ \\
 7  & $-(7.8088C_f+0.17466C_A)$     &  N  \\
 8  & $  2.4924C_f+1.4233C_A$       &  N  \\
 9  & $  2.3520C_f-1.0239C_A$       &  Y \ iff \ $r_C > 2.2970$ \\
10  & $-(1.7145C_f+0.059843C_A)$    &  Y  \\
11  & $-0.022326C_f+0.049590C_A$    &  Y \ iff \ $r_C < 0.450206$ \\
12  & $  0.37396C_f-0.27643C_A$     &  Y \ iff \ $r_C > 1.3528$  \\
13  & $-(0.108455C_f+0.21716C_A)$   &  Y  \\
14  & $-(0.025088C_f+0.25622C_A)$   &  Y  \\
15  & $ 0.020247C_f-0.33847C_A$     &  Y \ iff \ $r_C > 0.059818$  \\
16  & $-(0.0022434C_f+0.41598C_A)$  &  Y  \\
17  & $-(0.0013959C_f+0.51918C_A)$  &  Y  \\
18  & $0.00049000C_f-0.652215C_A$  &   Y \ iff \ $r_C > 0.75129 \times
10^{-3}$     \\
\hline\hline
\end{tabular}
\end{center}
\label{bellnfmax_nagt} 
\end{table}


\begin{thebibliography}{99}

\bibitem{rg}
%
Some early studies on this and related renormalization-group functions include
E. C. G. Stueckelberg and A. Peterman, Helv. Phys. Acta {\bf 26}, 499 (1953);
M. Gell-Mann and F. Low, Phys. Rev. {\bf 95}, 1300 (1954);
N. N. Bogolubov and D. V. Shirkov, Doklad. Akad. Nauk SSSR {\bf 103}, 391
(1955); C. G. Callan, Phys. Rev. D {\bf 2}, 1541 (1970);
K. Symanzik, Commun. Math. Phys. {\bf 18}, 227 (1970). See also K. Wilson,
Phys. Rev. D {\bf 3}, 1818 (1971).

\bibitem{sch2}
R. Shrock, Phys. Rev. D {\bf 88}, 036003 (2013) [arXiv:1305.6524].

\bibitem{sch}
T. A. Ryttov and R. Shrock, 
Phys. Rev. D {\bf 86}, 065032 (2012); Phys. Rev. D {\bf 86}, 085005 (2012). 

\bibitem{nlsm}
%
W. A. Bardeen, B. W. Lee, and R. E. Shrock, Phys. Rev. D {\bf 14}, 985 (1976);
E. Br\'ezin and J. Zinn-Justin, Phys. Rev. B {\bf 14}, 3110 (1976); see also 
A. Polyakov, Phys. Lett. B {\bf 59}, 79 (1975). 

\bibitem{landau}
L. Landau, in W. Pauli, ed., {\it N. Bohr and the Development of Physics}
(Pergamon, London, 1955). 

\bibitem{b2u1}
R. Jost and J. M. Luttinger, Helv. Phys. Acta {\bf 23}, 201 (1950).

\bibitem{b3u1Nf1}
E. de Rafael and J. L. Rosner, Annals of Phys. {\bf 82}, 369 (1974).

\bibitem{b3u1}
A. A. Vladimirov, Theor. Math. Phys. {\bf 43}, 417 (1980);
K. G. Chetyrkin, A. L. Kataev, and F. V. Tkachov, Nucl. Phys. B {\bf 174}, 345
(1980).

\bibitem{b4u1}
S. G. Gorishny, A. L. Kataev, S. A. Larin, and L. R. Surguladze, Phys. Lett. B
{\bf 256}, 81 (1991).

\bibitem{b5u1a}
A. L. Kataev and S. A. Larin, Pisma Zh. Eksp. Teor. Fiz. {\bf 96}, 64 
(2012) [JETP Lett. {\bf 96}, 61 (2012)]. 

\bibitem{b5u1b}
P. A. Baikov, K. G. Chetyrkin, J. H. K\"uhn, and J. Rittinger, JHEP 07, 017
(2012); 
P. A. Baikov, K. G. Chetyrkin, J. H. K\"uhn, and C. Sturm, Nucl. Phys. B
{\bf 867}, 182 (2013). 

\bibitem{jbw}
K. Johnson, R. Willey, and M. Baker, Phys. Rev. {\bf 163}, 1699 (1967).

\bibitem{adler}
S. L. Adler, Phys. Rev. D {\bf 5}, 3021 (1972).

\bibitem{scgt90}
%
For a review, see, e.g., the talks in the {\it 1990 International Workshop on
Strong Coupling Gauge Theories and Beyond, SCGT90}, eds. T. Muta and
K. Yamawaki (World Scientific, Singapore, 1991).

\bibitem{schierholz}
%
M. G\"ockeler, R. Horsley, E. Laermann, P. Rakow (DESY),
G. Schierholz, R. Sommer, U.-J. Wiese, Nucl. Phys. B {\bf 334}, 527 (1990);
M. G\"ockeler, R. Horsley, P. Rakow, and G. Schierholz, Phys. Rev. D {\bf 53},
1508 (1996); M. G\"ockeler, R. Horsley, V. Linke, P. Rakow, G. Schierholz, and
H. St\"uben, Phys. Rev. Lett. {\bf 80}, 4119 (1998).

\bibitem{kogut} A. Koci\'c, J. B. Kogut, and M.-P. Lombardo, Nucl. Phys. B {\bf
398}, 376 (1993); S. Kim, J. B. Kogut, and M.-P. Lombardo, Phys. Rev. D {\bf
65}, 054015 (2002); J. B. Kogut and C. G. Strouthos, Phys. Rev. D {\bf 71},
094012 (2005).

\bibitem{gies04}
H. Gies and J. Jaeckel, Phys. Rev. Lett. {\bf 93}, 110405 (2004). 

\bibitem{pascual} 
%
A. Palanques-Mestre and P. Pascual, Commun. Math. Phys. {\bf 95}, 277 (1984).
These authors' variable $K$ is equal to $4y$ and their coefficient 
$\beta_n$ is equal to $b_{n+2,n+1}/4^{n+1}$ in 
terms of the notation used here.

\bibitem{graceybeta}
J. A. Gracey, Phys. Lett. B {\bf 373}, 178 (1996).

\bibitem{graceygamma}
J. A. Gracey, Phys. Lett. B {\bf 317}, 415 (1993); 
M. Ciuchini, S. E. Derkachov, J. A. Gracey, and A. N. Manashov, 
Phys. Lett. B {\bf 458}, 117 (1999).

\bibitem{holdom2010}
%
B. Holdom, Phys. Lett. B {\bf 694}, 74 (2010). Note that the variable $A$ in
this paper is equal to $4y$ in terms of the notation used here.

\bibitem{ps}
C. Pica and F. Sannino, Phys. Rev. D {\bf 83}, 035013 (2011). 

\bibitem{b1}
D. J. Gross and F. Wilczek, Phys. Rev. Lett. {\bf 30}, 1343 (1973); 
H. D. Politzer, Phys. Rev. Lett. {\bf 30}, 1346 (1973); G. 't Hooft,
unpublished. 

\bibitem{gw1}
D. J. Gross and F. Wilczek, Phys. Rev. D {\bf 8}, 3633 (1973). 

\bibitem{b2}
W. E. Caswell, Phys. Rev. Lett. {\bf 33}, 244 (1974); 
D. R. T. Jones, Nucl. Phys. B {\bf 75}, 531 (1974). 

\bibitem{b3}
O. V. Tarasov, A. A. Vladimirov, and A. Yu. Zharkov, Phys. Lett. B {\bf 93},
429 (1980); S. A. Larin and J. A. M. Vermaseren, Phys. Lett. B {\bf 303}, 334
(1993). 

\bibitem{b4}
T. van Ritbergen, J. A. M. Vermaseren, and S. A. Larin, Phys. Lett. B {\bf
400}, 379 (1997). 

\bibitem{gross75}
D. J. Gross, in R. Balian and J. Zinn-Justin, eds. {\it Methods in Field
Theory}, Les Houches 1975 (North Holland, Amsterdam, 1976), p. 141.

\bibitem{ms}
G. 't Hooft, Nucl. Phys. B {\bf 61}, 455 (1973).

\bibitem{msbar}
W. A. Bardeen, A. J. Buras, D. W. Duke, and T. Muta, Phys. Rev. D {\bf 18},
3998 (1978).

\bibitem{bethke}
%
A recent review is S. Bethke, Eur. Phys. J. C {\bf 64}, 689 (2009).

\bibitem{bz}
T. Banks and A. Zaks, Nucl. Phys. B {\bf 196}, 189 (1982).

\bibitem{gk98}
E. Gardi and M. Karliner, Nucl. Phys. B {\bf 529}, 383 (1998);
E. Gardi, G. Grunberg and M. Karliner, JHEP {\bf 9807}, 007 (1998).

\bibitem{bvh}
T. A. Ryttov and R. Shrock, Phys. Rev. D {\bf 83}, 056011 (2011);
Phys. Rev. D {\bf 85}, 076009 (2012). 

\bibitem{bc}
R. Shrock, Phys. Rev. D {\bf 87}, 105005 (2013).

\bibitem{lnn}
R. Shrock, Phys. Rev. D {\bf 87}, 116007 (2013). 

\bibitem{rc}
T. A. Ryttov, arXiv:1309.3867, arXiv:1311.0848. 

\bibitem{gracey4loop}
J. A. Gracey, J. Phys. A {\bf 46}, 225403 (2013). 

\bibitem{pqcd}
%
There have been many studies of scheme transformations applicable in the
vicinity of the origin, $\alpha=0$, designed to optimize the accuracy and 
stability of perturbative QCD calculations; we do not consider these here,
since our focus is on a possible UV zero away from the origin. 

\bibitem{signconvention}
%
Since our previous related works \cite{bvh,sch,sch2,bc,lnn} focused on
asymptotically free non-Abelian gauge theories, it was convenient to extract an
overall minus sign multiplying the beta function. Because we focus here on
infrared-free theories, we write the beta function as in Eqs. (\ref{beta}) and
(\ref{beta_nagt}). 

\bibitem{nonan}
%
One can generalize this to scheme transformations with $f(a')$ functions that
have an essential zero at $a'=0$, as discussed in footnote [25] of the first
paper of \cite{sch}. The results for $b_\ell'$ are the same for this case as
for a transformation function $f(a')$ that is analytic at $a'=0$.

\bibitem{cgtu1}
%
In passing, we note that this result, that $b_1$ and $b_2$ have the same sign,
so the theory has no two-loop UV zero, applies more generally to a vectorial
U(1) gauge theory with $N_{f_i}$ fermions of different charges $q_i$ for
$i=1,...,k$, and, even more generally, to a chiral U(1) gauge theory, with
left-handed fermions $\psi_{i,L}$, $1 \le i \le N_{f,L}$ and charges $q_{iL}$,
and right-handed fermions $\psi_{i,R}$, $1 \le i \le N_{f,R}$ and charges
$q_{iR}$.  In the chiral case, one requires that $\sum_{i=1}^{N_{f,L}}
q_{i,L}^3 - \sum_{j=1}^{N_{f,R}} q_{i,R}^3 = 0$ so that there is no chiral
gauge anomaly.

\bibitem{nfintegral}
%
Here and below, when an expression is given for $N_f$ that formally evaluates
to a non-integral real value, it is understood implicitly that one infers
an appropriate integral value of $N_f$ from this. 

\bibitem{holdif}
%
Our expansions agree with, and extend to higher order, the results in
\cite{holdom2010} except for the $A^4$ term in Eq. (5) of \cite{holdom2010};
that term was derived from an earlier partial calculation of $b_5$ in
P. A. Baikov, K. G. Chetyrkin, and J. H. K\"uhn, Phys. Rev. Lett.  {\bf 88},
012001 (2001) which was superseded by the full calculation in \cite{b5u1b}.

\bibitem{casimir}
%
Our notation for Casimir invariants is $C_2(G) \equiv C_A$ and, for a fermion
representation $R$, $C_2(R) \equiv C_f$ and $T(R) = T_f$; our normalizations
are standard, so that, e.g., for $G={\rm SU}(N_c)$, $C_A=N_c$, and, for $R$
equal to the fundamental representation, $C_f = (N_c^2-1)/(2N_c)$ and
$T_f=1/2$.

\bibitem{zeta} 
I. S. Gradshteyn and I. M. Ryzhik, {\it Tables of
Integrals, Series, and Products} (Academic Press, New York, 1980),
eqs. (8.321); E. C. Titchmarsh, {\it Theory of the Riemann Zeta Function}
(Oxford Press, Oxford, 1951).

\end{thebibliography}
\end{document}